\documentclass[sigconf]{acmart}

\usepackage{booktabs} 

\usepackage{tabularx} 

\usepackage{setspace}
\usepackage{placeins}
\usepackage{afterpage}
\usepackage[utf8]{inputenc}
\usepackage{graphicx}
\usepackage{balance}
\usepackage{amsthm}
\usepackage{color}
\usepackage{caption}
\usepackage{amsmath}
\usepackage{tabu}
\usepackage{array}
\usepackage{booktabs}
\usepackage{threeparttable}
\usepackage{relsize}
\usepackage{physics}
\usepackage{algorithm}
\usepackage[noend]{algpseudocode}

\newlength{\maxwidth}
\newcommand{\algalign}[2]
{\makebox[\maxwidth][r]{$#1{}$}${}#2$}

\usepackage{float}
\usepackage{stfloats}
\usepackage{lipsum}
\usepackage[english]{babel}

\usepackage{url}

\usepackage[T1]{fontenc}
\usepackage{mwe}    
\usepackage{subfig}
\let\oldnl\nl
\newcommand{\nonl}{\renewcommand{\nl}{\let\nl\oldnl}}
\newtheorem{thm}{Theorem}
\newtheorem{problem}{Problem}

\theoremstyle{definition}
\newtheorem{defn}{Definition}

\def\Equal{\texttt{=}}

\AtBeginDocument{%
  }

\setcopyright{acmlicensed}
\copyrightyear{2018}
\acmYear{2018}
\acmDOI{XXXXXXX.XXXXXXX}

\acmConference[EDBT]{sshaham@usc.edu}{March 25--28,
 2025}{Barcelona}
\acmISBN{978-1-4503-XXXX-X/18/06}

\begin{document}
\settopmatter{printacmref=false}

\title{Differentially Private Publication of Electricity Time Series Data in Smart Grids}



\author{Sina Shaham}
\affiliation{%
  \institution{University of Southern California}
  \city{Los Angeles}
  \country{USA}}
\email{sshaham@usc.edu}

\author{Gabriel Ghinita}
\affiliation{%
  \institution{Hamad Bin Khalifa University}
  \city{Ar-Rayyan}
  \country{Qatar}}
\email{gghinita@hbku.edu.qa}

\author{Bhaskar Krishnamachari}
\affiliation{%
  \institution{University of Southern California}
  \city{Los Angeles}
  \country{USA}}
\email{bkrishna@usc.edu}

\author{Cyrus Shahabi}
\affiliation{%
  \institution{University of Southern California}
  \city{Los Angeles}
  \country{USA}}
\email{shahabi@usc.edu}



\begin{abstract}
Smart grids are a valuable data source to study consumer behavior and guide energy policy decisions. In particular, time-series of power consumption over geographical areas are essential in deciding the optimal placement of expensive resources (e.g.,  transformers, storage elements) and their activation schedules. However, publication of such data raises significant privacy issues, as it may reveal sensitive details about personal habits and lifestyles. Differential privacy (DP) is well-suited for sanitization of individual data, but current DP techniques for time series lead to significant loss in utility, due to the existence of temporal correlation between data readings. We introduce {\em STPT (Spatio-Temporal Private Timeseries)}, a novel method for DP-compliant publication of electricity consumption data that analyzes spatio-temporal attributes and captures both micro and macro patterns by leveraging RNNs. Additionally, it employs a partitioning method for releasing electricity consumption time series based on identified patterns. We demonstrate through extensive experiments, on both real-world and synthetic datasets, that STPT significantly outperforms existing benchmarks, providing a well-balanced trade-off between data utility and user privacy. 

\end{abstract}


\maketitle


\section{Introduction}

Analysis of electricity consumption data plays a critical role in planning power grid infrastructures for smart cities. Such data are represented as time series of geo-tagged data, and serve as a critical information source, providing detailed knowledge to policymakers about energy usage trends. Such data can help identify where/when consumption surges occur, and decide where to place expensive equipment such as transformers; or in the case of renewable energy sources, it can help decide where to place storage elements, and when to release their capacity back into the grid.

Despite its advantages, analysis of electricity consumption time series raises significant privacy concerns. 
The data may reveal personal habits and lifestyles, such as individuals' daily routines, working hours, etc., leading to privacy violations.  Moreover, the risk of third-party exploitation by marketers and advertisers poses a threat of unwanted privacy intrusions, as consumers may be targeted based on their specific energy usage behavior. 

Prevailing approaches for protecting electricity time series information rely on the powerful Differential Privacy (DP) model~\cite{dwork2008differential}. DP achieves privacy by adding noise to the data, thereby minimizing the likelihood of re-identification. 
However, when dealing with time series data, the existence of temporal correlations leads to increased re-identification risk over time, causing DP to add excessive noise to offset this risk, lowering data utility~\cite{fan2013adaptive,leukam2021privacy}.



We propose a model that jointly takes into account both time and space attributes of electricity consumption data. Our Spatio-Temporal Private Timeseries (STPT) algorithm  trains a Recurrent Neural Network (RNN) to identify spatio-temporal electricity consumption patterns. The patterns are subsequently used to partition the time series into a spatio-temporal histogram that is used by DP mechanisms to sanitize and release the data. 
A key innovation of STPT is 
the incorporation of spatial distribution alongside temporal sequencing. We start with a low-granularity aggregation of time series data to identify macro consumption trends, followed by several increasingly-higher granularity aggregations to discern micro trends. This dual-focus on both macro and micro trends allows for a nuanced representation of consumption patterns, enhancing STPT's ability to preserve data utility while enforcing DP.
Note that, while RNNs have been used before for geo-tagged time series, the focus of prior work is on trajectory forecasting~\cite{shahabiRNN}. Our approach is significantly different as consumer locations are static (e.g., households), and the purpose of RNN is to estimate future power consumption at each location.

Our specific contributions include:

\begin{itemize}
    \item We introduce a novel method for modeling and representing electricity time series data, which takes into account both spatial and temporal properties.
    \item We propose STPT, an innovative algorithm that integrates a unique approach for training RNNs across both time and space dimensions on differentially private data. 
    \item We design a customized technique for STPT that clusters electricity consumption data across time and space, thereby improving data utility when applying DP-compliant mechanisms for sanitization of time series. 
    \item We perform an extensive experimental evaluation on both real-world and synthetic datasets, demonstrating STPT's notable improvements in data utility compared to existing benchmarks \footnote{Codes and datasets are publicly available online at the following link: https://github.com/ANRGUSC/pars/}. 
\end{itemize}

Section~\ref{section: preliminaries} introduces foundational concepts and the system architecture. Section~\ref{section: spatio-temporal time series} outlines our method for electricity data modeling, following by the description of the STPT algorithm in Section~\ref{section: STPT}. We provide an extensive experimental evaluation in comparison with several benchmarks in Section~\ref{section: experimental evaluation}. 
Section~\ref{Sec: related work} reviews related work. Section~\ref{section: conclusion} concludes with future work directions. Proofs of theorems are provided in the appendix.

\section{Preliminaries}\label{section: preliminaries}
Consider a two-dimensional map that encloses a set of $N$ households $\mathcal{U} = \{ u_1,...,u_N \}$. We denote the electricity consumption for user $i$ at time $t$ by $x_{i,t}$ (we use the term household and power grid user interchangeably). Each household meter sends its electricity reading to an aggregator at regular intervals $\Delta \times t $ ($t=1,...,T$) where $\Delta \in \mathbb{R}$. The  dataset of meter readings is denoted as: 

\begin{equation}
    \mathcal{D} = (x_{i,t})_{i=1,...,N; t=1,...,T}
\end{equation}

The goal is to release the dataset $\mathcal{D}$ according to the requirements of DP, thus preventing an adversary from inferring the consumption patterns of any individual user. We start our discussion by explaining the system model commonly used for the publication of DP electricity consumption time series, followed by an illustration of the foundational concepts related to DP. A summary of notations used throughout the paper is provided in Appendix~\ref{appendix: table of notations}.

\subsection{System Model}
\begin{figure}[t]
\includegraphics[scale=.5]{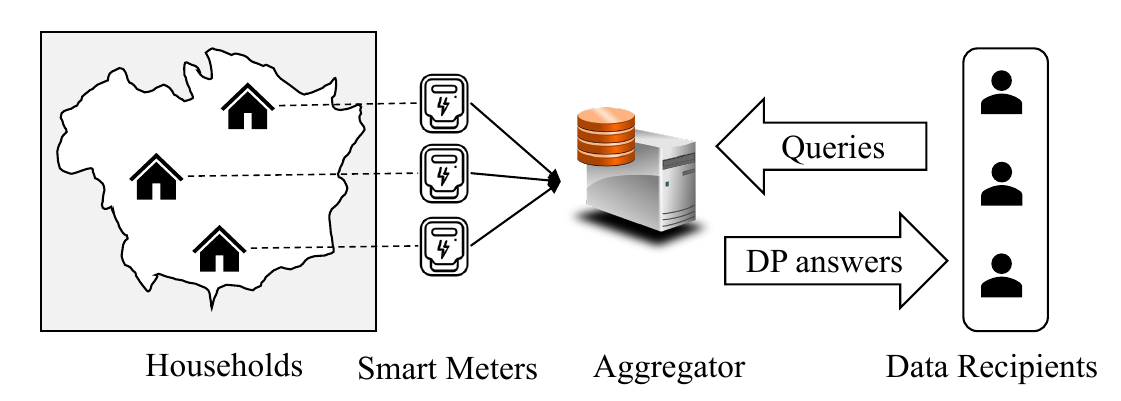}
\centering
\caption{System model}
\label{Fig: system model}
\end{figure}

Figure~\ref{Fig: system model} depicts the system architecture which adheres to the industry-standard model for publishing electricity data, and consists of households, a data aggregator, and data recipients which perform analyses on the released consumption data. The specific functions of each party are detailed below.
\begin{itemize}
    \item {\em Households} equipped with smart meters are generators of data and are considered to be trustworthy in the system model. The electricity consumption of users is recorded hourly using their meter and sent to the data aggregator. 
    \item {\em Data Aggregator} is a trusted party that collects the time series generated by users and publishes their aggregated data in a privacy-preserving way. The sanitization process is done based on DP, preventing adversaries from inferring any individual-level consumption pattern. 
    \item {\em Data Recipients} leverage the private aggregated data for diverse applications, from forecasting to planning. Their objective is to utilize consumption values over specific spatial regions and time periods. The recipients are considered to be honest but curious, and they may attempt to infer individual user consumption from aggregated data. Individual consumption details must be protected, as they can be used to infer sensitive details about users, such as activity patterns, lifestyle habits, etc.
\end{itemize}

\subsection{Differential Privacy}\label{Sec: differential privacy}

Two databases $\mathcal{D}$ and $\mathcal{D}'$ are called  {\em neighboring} or {\em sibling} if they differ in a single record $t$, i.e., $\mathcal{D}'=\mathcal{D}\bigcup \{ t\}$ or $\mathcal{D}'=\mathcal{D}\text{\textbackslash} \{ t\}$. 

\begin{defn}[$\epsilon$-Differential Privacy\cite{dwork2008differential}] 
    A randomized mechanism $\mathcal{A}$ provides $\epsilon$-DP if for any pair of neighbor datasets $D$ and $D'$, and any $a\in Range(\mathcal{A})$,
    \begin{equation}
        \dfrac{Pr(\mathcal{A}(\mathcal{D})=a)}{Pr(\mathcal{A}(\mathcal{D}')=a)} \leq e^\epsilon 
    \end{equation}
    
\end{defn}

Parameter $\epsilon$ is referred to as privacy budget.
$\epsilon$-DP requires that the output obtained by executing mechanism $\mathcal{A}$ does not significantly change by adding or removing one record in the database. Thus, an adversary is not able to infer with significant probability whether an individual's record was included or not in the database.

Aside from the amount of privacy budget, another factor that plays a critical role in achieving $\epsilon$-DP is the concept of {\em sensitivity}, which captures the maximal difference achieved in the output by adding or removing a single record from the database.

\begin{defn}[$L_1$-Sensitivity\cite{dwork2006calibrating}]
Given sibling datasets $\mathcal{D}$, $\mathcal{D}'$ the $L_1$-sensitivity of a set $g = \{g_1, \ldots, g_m\}$ of real-valued functions is:
\begin{equation}
s  =  \underset{\forall \mathcal{D},\mathcal{D}'}{max} \; \sum_{i\Equal1}^m |g_i(\mathcal{D})- g_i(\mathcal{D}')|
\end{equation}

\end{defn}

A widely-used mechanism to achieve $\epsilon$-DP is called Laplace mechanism. This approach adds to the output of every query function noise drawn from Laplace distribution Lap$(b)$ with scale $b$ and mean $0$, where $b$ depends on sensitivity and privacy budget.

\begin{equation}
    \text{Lap}(x|b) = \dfrac{1}{2b}e^{-|x|/b}\; \text{where }\; b=\dfrac{s}{\epsilon} 
\end{equation}
To simplify notation, we denote Laplace noise by $\text{Lap}( \dfrac {s } {\epsilon} )$.


In our work, we make extensive use of the following three essential results in differential privacy:

\begin{thm}[Sequential Composition \cite{mcsherry2009privacy}]\label{thm: sequential}
Let $A_1$ and $A_2$ be two DP mechanisms that provide $\epsilon_1$- and $\epsilon_2$-differential privacy, respectively. Then, applying in sequence $A_1$ and $A_2$ over the dataset $\mathcal{D}$ achieves $(\epsilon_1+\epsilon_2)$-differential privacy.
\end{thm}

\begin{thm}[Parallel Composition \cite{mcsherry2009privacy}]\label{thm: parallel}
Let $A_1$ and $A_2$ be two DP mechanisms that provide $\epsilon_1$- and $\epsilon_2$-differential privacy, respectively. Then, applying $A_1$ and $A_2$ over two {\em disjoint} partitions of the dataset $\mathcal{D}_1$ and $\mathcal{D}_2$ achieves $(\max{(\epsilon_1,\epsilon_2}))$-differential privacy.
\end{thm}

\begin{thm}[Post-Processing Immunity\cite{dwork2008differential}]\label{thm: immunity}
Let \( A \) be an \( \varepsilon \)-differentially private mechanism and \( g \) be
an arbitrary mapping from the set of possible output sequences \( O \) to an arbitrary set. Then, \( g \circ A \) is \( \varepsilon \)-differentially private.
\end{thm}

\section{Problem Statement}
\label{section: spatio-temporal time series}


\subsection{Time-Series Representation}

Consider a spatial grid of size $C_x \times C_y$ overlaid on a 2D map, dividing the spatial domain into smaller regions. Additionally, we divide the time dimension into a number of $C_t$ equal-length intervals. The electricity consumption data is thus captured by a three-dimensional matrix $\mathcal{C}_{\text{cons}}$ called consumption matrix with $C_x \times C_y \times C_t$ elements. 
Each element $c_{ijk}$ in this matrix represents the electricity consumption within the $(i,j)$ region during the time interval from $\Delta \times k$ to $\Delta \times (k+1)$, where $\Delta$ is the time resolution. 
For ease of analysis, especially when conducting sensitivity studies in relation to data publication under DP, we assume without loss of generality that $\Delta = 1$. This assumption implies that each data point in the time series corresponds to distinct time intervals, meaning that $C_t$ is effectively the length of the time series $(C_t = T)$.

\subsection{Problem Formulation}\label{section: Problem Formulation}

Data recipients are interested in answering multi-dimensional {\em range queries} on top of the electricity consumption matrix. 
 
\begin{defn} {\em (Range Query)}
A range query on the consumption matrix is a $3$-orthotope with dimensions denoted as $d_1\times d_2 \times d_3$, where $d_i$ represents a continuous interval in dimension $i$. 
\end{defn}

To evaluate the accuracy of results, we use the {\em Mean Relative Error (MRE)} metric.
For a query $q$ with the true aggregated consumption $p$ and noisy consumption value $\overline{p}$, MRE is calculated as 
\begin{equation}\label{Equation: MRE}
   MRE(q) =  \dfrac{|p - \overline{p}|}{p}\times 100
\end{equation}


\begin{problem}\label{problem statement}
Given a consumption matrix denoted by $\mathcal{C}_{\text{cons}}$, generate a $\epsilon$-DP matrix $\mathcal{C}_{\text{sanitized}}$ such that average MRE subject to range queries is minimized. 
\end{problem}

\begin{figure}[t]
    \centering 
    \subfloat[Electricity consumption matrix.]{%
        \includegraphics[scale=.23]{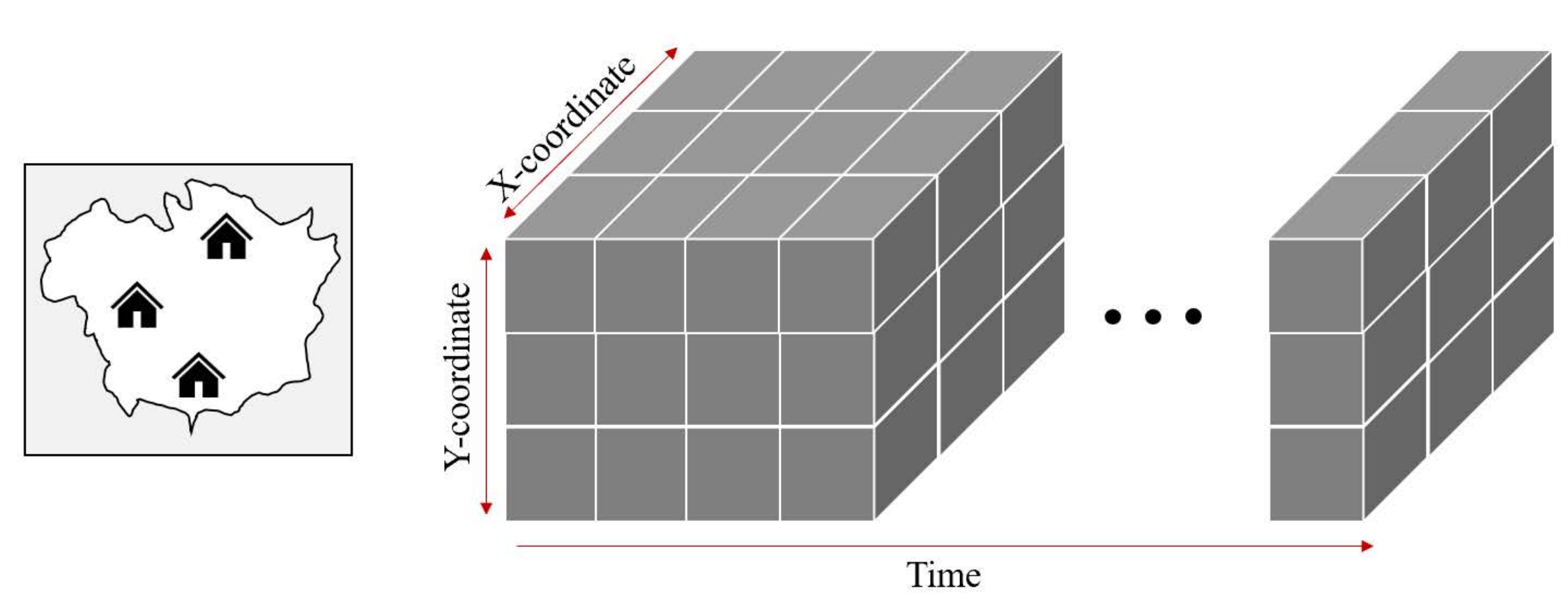}
        \label{Fig: consumption matrix - a}
    }
    \hfill 
    \subfloat[Generated time series for training the RNN unit.]{%
        \includegraphics[scale=.2]{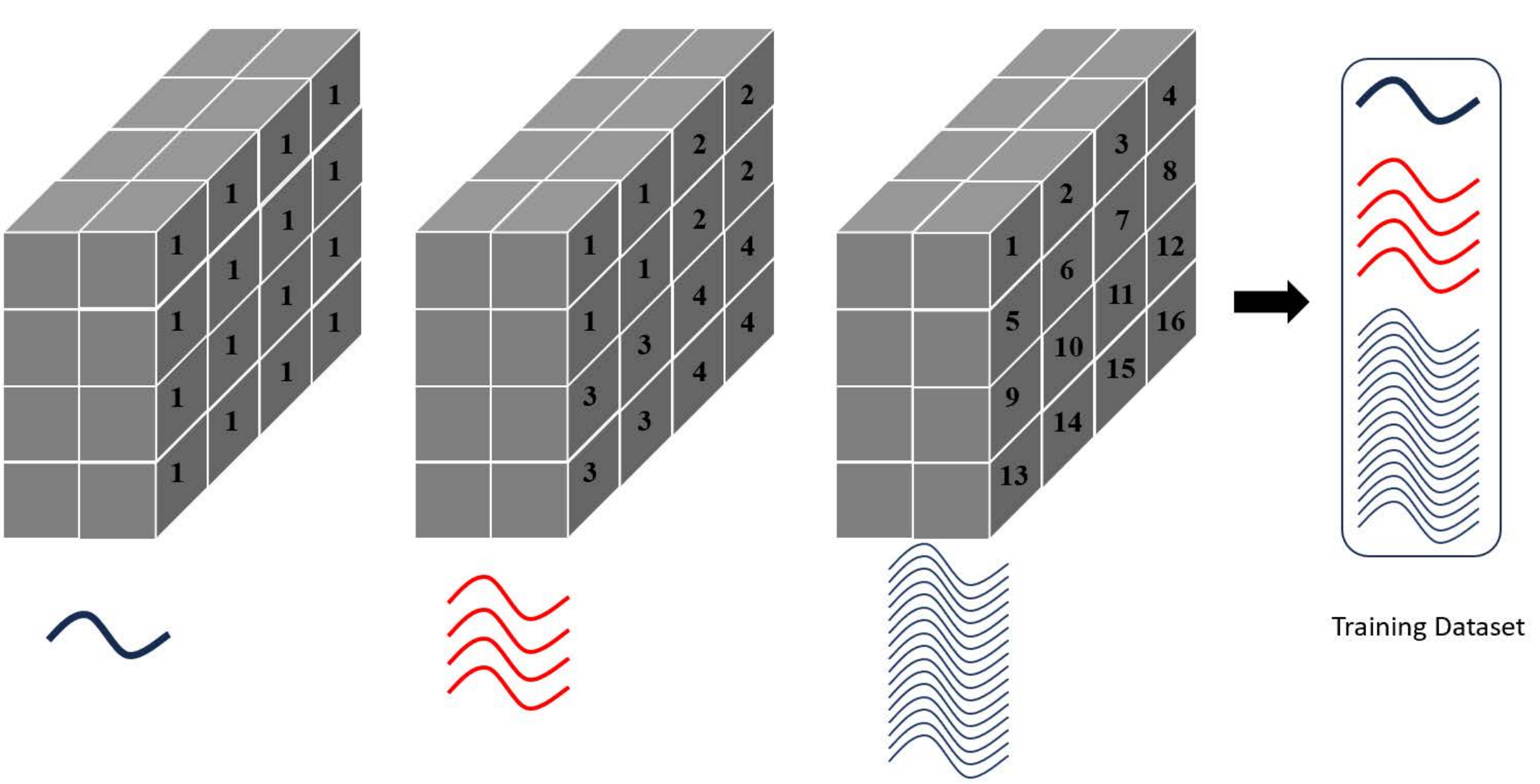}
        \label{Fig: consumption matrix - b}
    }
    \hfill 
    \subfloat[Clustering hypercube cells for sanitization purposes.]{%
        \includegraphics[scale=.3]{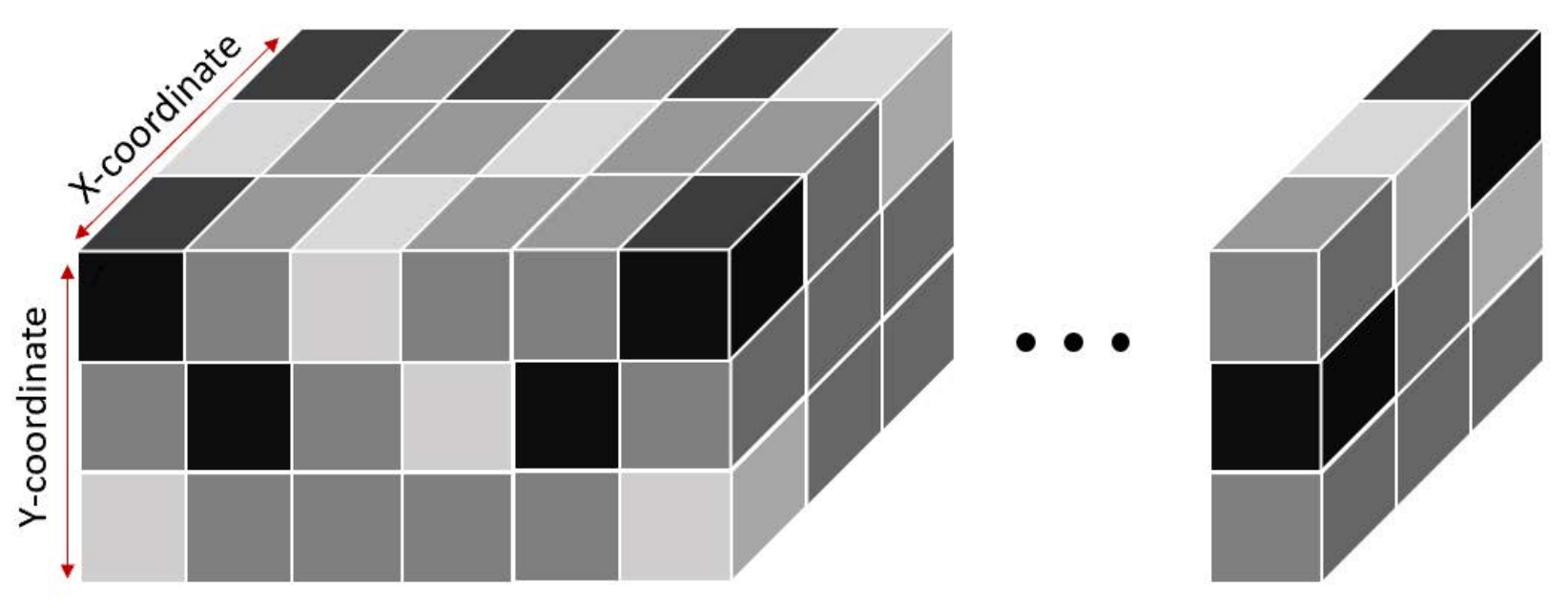}
        \label{Fig: consumption matrix - c}
    }
    \hfill 
    \caption{Consumption matrix in different stages.}
    \label{Fig: consumption matrix}
\end{figure}

\subsection{A Simple Strategy}\label{section: simple strategy}

One simple strategy to publish the electricity consumption matrix is the {\em Identity} algorithm~\cite{xu2013differentially}. This algorithm was initially designed for population histograms, and works by adding independent Laplace noise to every matrix cell. When applying this technique to the consumption matrix, it is essential to note that time series have temporal correlations. As a result, every snapshot of time should have its distinct allocated privacy budget, according to the sequential composition theorem (Theorem~\ref{thm: sequential}). Conversely, since at each timestamp the spatial grid creates disjoint partitions of the map, parallel composition applies within each time interval (Theorem~\ref{thm: parallel}).
The following important result emerges which quantifies the sensitivity of a query on each cell of the electricity consumption matrix.


\begin{thm}\label{theorem: sensitivity simple strategy}
    The sensitivity of range queries of size $1 \times 1 \times 1$ on the electricity consumption matrix $\mathcal{C}_{\text{cons}}$ is given by $\underset{i,t}{\max}\; x_{i,t}$.

\end{thm}

\begin{thm}\label{theorem: decomposition}
    The consumption matrix follows sequential decomposition in time and parallel decomposition in space. 
\end{thm}

The Identity algorithm allocates an equal amount of privacy budget to each time slice. Therefore, if $\epsilon_{\text{tot}}$ represents the entire budget allocated for sanitization, the budget for each time slice amounts to $\epsilon_{\text{tot}}/C_t$. Then, each cell of the matrix is sanitized by the addition of Laplace noise with sensitivity $1$ and privacy budget $\epsilon_{\text{tot}}/C_t$ given that the time series are normalized in advance.


\section{STPT Algorithm}\label{section: STPT}

\subsection{Overview}

The STPT algorithm starts by generating two matrices $\mathcal{C}_{\text{cons}}$ and $\mathcal{C}_{\text{norm}}$ out of the collected time series from different neighborhoods of the map. $\mathcal{C}_{\text{cons}}$ denotes the consumption matrix based on the actual values of the time series, whereas $\mathcal{C}_{\text{norm}}$  is its normalized counterpart. 
We employ min-max normalization at a global level. Under this scheme, the normalized consumption for user $i$ at time $j$ is computed as follows:
\begin{equation}
    x_{i,j} := \dfrac{x_{i,j} - \underset{i,t}{\text{min}}\, x_{i,t} }{ \underset{i,t}{\text{max}}\, x_{i,t}- \underset{i,t}{\text{min}}\, x_{i,t} }
\end{equation}

\begin{figure}[t]
\includegraphics[scale=.35]{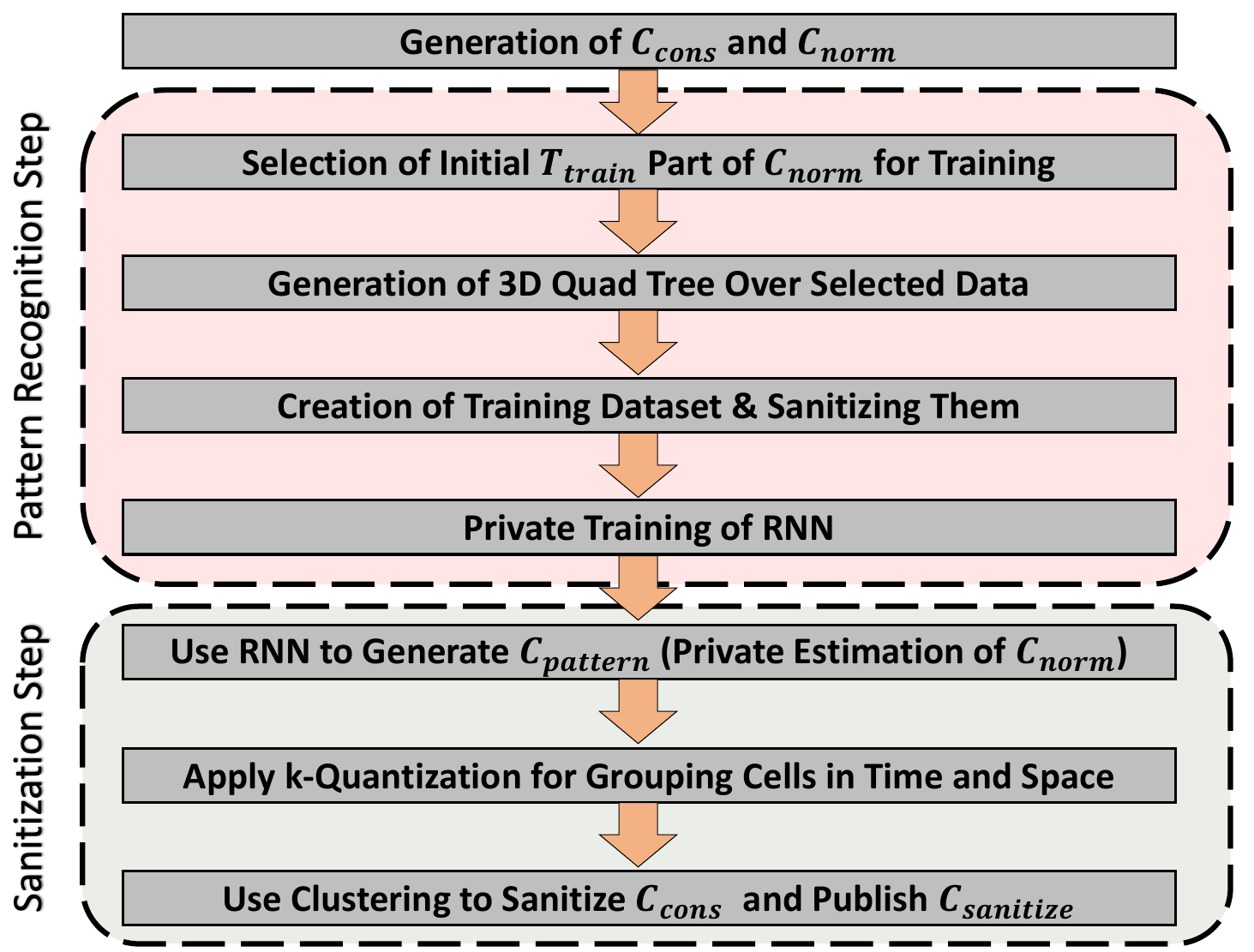}
\caption{Workflow of Proposed Approach}
\label{Fig: workflow}
\end{figure}

The STPT algorithm conducts two sequential core procedures to generate the DP consumption matrix, namely the Pattern Recognition Step, followed by the Sanitization Step. The workflow of the approach is shown in Figure~\ref{Fig: workflow}.

\noindent
{\bf Pattern Recognition Step.}
The primary aim of pattern recognition is to create a sanitized version $\mathcal{C}_{\text{pattern}}$ of the normalized consumption matrix $\mathcal{C}_{\text{norm}}$. The core objective is to achieve rough estimations of the normalized time series values while utilizing a small amount of privacy budget. The choice of using $\mathcal{C}_{\text{norm}}$ over $\mathcal{C}_{\text{cons}}$ is strategic, as it helps in bounding the sensitivity of the cells during the sanitization process. The generation of sanitized estimated values in $\mathcal{C}_{\text{pattern}}$ involves using a short segment of the time series, $T_{\text{train}}$, to predict future consumption while preserving privacy. The training data are sanitized through a novel hierarchical method, considering both time and space dimensions. The sanitized data are used to train a RNN, which is responsible for estimating the remaining values in $\mathcal{C}_{\text{pattern}}$. The total privacy budget allocated for this phase is denoted by $\epsilon_{\text{pattern}}$.

\noindent
{\bf Sanitization Step.}
This algorithm's primary objective is to perform an intelligent partitioning of the matrix, based on the private estimates in $\mathcal{C}_{\text{pattern}}$, and then to sanitize and release the values of $\mathcal{C}_{\text{cons}}$. Since the estimates in $\mathcal{C}_{\text{pattern}}$ are private, the resulting matrix partitioning is also privacy-preserving, being derived from private data. The partitioning approach for the consumption matrix, both temporally and spatially, is predicated on the principle of homogeneity. This principle, which contributes to enhanced data utility, aims to group cells with similar values into the same partition. Post partitioning, the true values in each partition, extracted from $\mathcal{C}_{\text{cons}}$, are aggregated and sanitized. The final output of this procedure is the matrix $\mathcal{C}_{\text{sanitized}}$, representing the differentially private version of $\mathcal{C}_{\text{cons}}$. The privacy budget for the sanitization algorithm is denoted as $\epsilon_{\text{sanitize}}$. This leads to the total amount of privacy budget of $\epsilon_{\text{tot}}$ for the STPT algorithm where,

\begin{equation}
    \epsilon_{\text{tot}} = \epsilon_{\text{sanitize}} + \epsilon_{\text{pattern}}
\end{equation}

Therefore, STPT publishes a $\epsilon_{\text{tot}}$-differential private version of the original consumption matrix ($\mathcal{C}_{\text{cons}}$). 
The pseudocode for STPT is provided in Algorithm~\ref{Algo: STPT}.

\subsection{Pattern Recognition}

The goal of the pattern recognition phase in the STPT algorithm is to effectively use a designated privacy budget, $\epsilon_{\text{pattern}}$, to develop a method for privately generating approximate estimates for cells within the normalized consumption matrix. The primary means to accomplish this is through the private training of an RNN unit. The input comprises of household time series data along with their corresponding geographic locations on the map. The pattern recognition process is outlined in Section~\ref{section: spatio-temporal time series}. It involves generating the consumption matrix $\mathcal{C}_{\text{cons}}$ from the time series data and creating $\mathcal{C}_{\text{norm}}$, the consumption matrix based on normalized time series.

Next, the initial time segment ($T_{\text{train}}$) of the consumption matrix $\mathcal{C}_{\text{norm}}$ is allocated for training, resulting in a matrix dimensionality of $C_x \times C_y \times C_t[0:T_{\text{train}}]$. The notation $C_t[0:T_{\text{train}}]$ indicates the selection of indices from $0$ to $T_{\text{train}}$ on the time dimension. A critical challenge is determining an efficient training method for the RNN, ensuring it comprehensively learns both micro and macro trends across neighborhoods, while minimizing the amount of privacy budget utilized for training. One straightforward training method for the RNN model involves the sanitization strategy described in Section~\ref{section: simple strategy}. By adopting this method, every time snapshot is allocated a budget of $\epsilon_{\text{sanitize}}/ T_{\text{train}}$, and each matrix entry undergoes Laplace noise perturbation with a sensitivity of one and budget $\epsilon_{\text{sanitize}}/ T_{\text{train}}$, which translates to $Lap(1/ (\epsilon_{\text{sanitize}}/ T_{\text{train}}))$ given that time series are normalized.

Despite its feasibility, this method introduces excessive noise into the training data, impacting model accuracy. We introduce an approach centered on the generation of a spatio-temporal quadtree (lines~\ref{line: quadtree start} to~\ref{line: quadtree end} in Algorithm~\ref{Algo: STPT}). Assuming $C_x< C_y$, the process initiates by segmenting time into $log_2 (C_x)+1$ levels, resulting in a time span of $T_{\text{train}}'$ for each interval, derived as follows:
\begin{equation}
   T_{\text{train}}'=  \lceil  T_{\text{train}}/(log_2 (C_x)+1) \rceil
\end{equation}

The matrix corresponding to the $i^{th}$ interval is $C_x \times C_y \times C_z[i*T_{\text{train}}':(i+1)*T_{\text{train}}']$, corresponding to the quadtree's $i^{th}$ level. In the first segment of the matrix corresponding to the tree's root, all cells are presumed to be part of a unified neighborhood. However, in the subsequent sub-matrix (depth $1$), the previous matrix's neighborhoods are subdivided into four distinct quadrants. Given that quadtrees are data-independent index structures, we do not need to expend privacy budget to determine the division points of the space. Once the spatio-temporal partitioning of the training matrix is completed, there exist $4^i$ neighborhoods at each level $i$. The next step of the algorithm is generating a single time series representing each neighborhood. The representative time series is generated by element-wise averaging of all time series in the neighborhood over the time allocated for that  level of the tree. Consider a neighborhood at the $i^{th}$ level of the tree, and without loss of generality, suppose times series $1,..,j$ fall in this neighborhood. For each point in time $t$ lying in the interval $i*T_{\text{train}}':(i+1)*T_{\text{train}}'$ the value in the representative time series is the average of all consumption of users in that neighborhood and that specific time, calculated as:
\begin{equation}\label{equation: averaging}
    x_{\text{rep},t} = \dfrac{1}{j}\sum_{i\Equal 1}^j x_{i,t}
\end{equation}

We emphasize that the time series created are stacked, and not sequential. To produce training data for subsequent training phases, a time window is swept across each time series individually to generate the training data. An illustration of the method is exemplified in Figure~\ref{Fig: consumption matrix - b}. As observed, we utilize a $4\times 4\times 6$ matrix for the training process. The entire duration of training is segmented into $3$ parts, which translates to a duration of $6/(log_2(4)+1)$ for each part. This involves the creation of $3$ submatrices, each having dimensions of $4\times 4 \times 2$. The root node of the quadtree includes only one neighborhood, indicated by the cells' number, which results in one distinct time series shown beneath. Depths $1$ and $2$ in the tree align with $4$ and $16$ time series, respectively. Altogether, $21$ time series are employed for the next step.

Once partitioning is complete, the $4^i$ resulting time series at each level $i$  undergo sanitization. The key advantage of hierarchical partitioning lies in bounding sensitivity. As outlined in Theorem~\ref{theorem: sensitivity}, the sensitivity of the time series at depth $i$ is  $1/ 4^{log_2(C_x)-i}$. The underlying principle suggests that macro trends can be captured with heightened precision since the sensitivity of the time series is reduced, allowing for a smaller amount of Laplace noise during sanitization. The stacked sanitized time series corresponds to list $res$ in Algorithm~\ref{Algo: STPT}. After finalizing the time series, training data for the RNN is produced by sweeping a time window across the time series and organizing them into batches. This training data is then employed for training an RNN. Subsequently, the RNN is utilized to create private estimations of the matrix $\mathcal{C}_{\text{norm}}$ in $\mathcal{C}_{\text{pattern}}$.

\begin{thm}\label{theorem: sensitivity}
    The sensitivity of a cell at depth $i$ of the matrix is $1/ 4^{log_2(C_x)-i}$.
\end{thm}

\begin{algorithm}[t]
\caption{STPT}
\label{Algo: STPT}
\begin{algorithmic}[1]
\Statex \textbf{Input:} $\mathcal{D}$, $\epsilon_{\text{pattern}}$, $\epsilon_{\text{sanitize}}$, Quad Tree Depth ($depth$), Window Size ($ws$), Training Time ($T_{\text{train}}$), Quantization Level ($k$).
\Statex \textbf{Output:} Sanitized Consumption Matrix
\State $\mathcal{C}_{\text{cons}}\leftarrow$ Create Consumption Matrix
\State $\mathcal{C}_{\text{norm}}\leftarrow$ Min-Max Normalize $\mathcal{C}_{\text{cons}}$
\State Select $T_{\text{train}}$ Data Points from $\mathcal{C}_{\text{norm}}$
\State $res\leftarrow []$ \Comment{Initialize empty list for time series}
\For{$d \in [0, \dots, depth]$} \label{line: quadtree start}
    \State $Temp\leftarrow$ Select time interval $[i \cdot T_{\text{train}}:(i+1) \cdot T_{\text{train}}]$
    \State Divide $x$ and $y$ Axes of $Temp$ into $2^d$ Creating $4^d$ Neighborhoods
    \For{each neighborhood}
        \State Compute Representative Time Series (Eq.~\ref{equation: averaging})
        \State Sanitize Time Series with Budget $\epsilon_{\text{pattern}}/T_{\text{train}}$ and Sensitivity $1/4^{log_2(C_x) - i}$
        \State Append Sanitized Series to $res$
    \EndFor
\EndFor
\State Prepare Training Data from $res$ Based on $ws$\label{line: quadtree end}
\State Train RNN
\State Generate $\mathcal{C}_{\text{pattern}}$ Using RNN
\State $\mathcal{P}\leftarrow k$-Quantize $\mathcal{C}_{\text{pattern}}$\label{line: sanitize start}
\For{each partition $P_i \in \mathcal{P}$}
    \State $f(P_i)\leftarrow$ Sum Values in $\mathcal{C}_{\text{cons}}$ for $P_i$
    \State $s\leftarrow$ Compute Sensitivity of $P_i$
    \State Sanitize $f(P_i)$ Using $s$ and Budget (Eq.~\ref{equation: budget})
    \For{each cell $c \in P_i$}
        \State Update $c$ in $\mathcal{C}_{\text{cons}}$ to $f(P_i)/|P_i|$
    \EndFor
\EndFor
\State \textbf{return} Sanitized $\mathcal{C}_{\text{cons}}$, i.e., $\mathcal{C}_{\text{sanitized}}$ \label{line: sanitize end}
\end{algorithmic}
\end{algorithm}


\subsection{Sanitization Algorithm}

The output of pattern recognition is the matrix $\mathcal{C}_{\text{pattern}}$, with dimensions $C_x \times C_y \times C_t$. Each element of this matrix is created using a differentially private approach. These elements are sanitized estimates of normalized time series, providing an idea of consumption patterns rather than actual consumption amounts. The purpose of the sanitization algorithm is not only to reveal these patterns but also to provide sanitized consumption values.

The sanitization algorithm of STPT (lines~\ref{line: sanitize start} to~\ref{line: sanitize end} in Algorithm~\ref{Algo: STPT}) starts by developing a non-overlapping partitioning of the matrix $\mathcal{C}_{\text{pattern}}$. The developed partitioning's objective is to group cells with similar values together. For this purpose, we use a $k$-quantization of matrix $\mathcal{C}_{\text{pattern}}$ to generate clusters. The formal definition of $k$-quantization is provided in Definition~\ref{definition: quantization}.

\begin{defn}[$k$-Quantization]\label{definition: quantization}
Let $\mathcal{C}_{\text{pattern}}$ be a 3-dimensional matrix with elements $c_{i,j,k}$, where $i, j, k$ are the indices of the matrix, and let $k$ be a positive integer representing the number of quantization levels. The $k$-quantization of $\mathcal{C}_{\text{pattern}}$ is a process defined as follows:

\begin{enumerate}
    \item \textbf{Determine Range:} Identify the minimum $\min(\mathcal{C}_{\text{pattern}})$ and maximum $\max(\mathcal{C}_{\text{pattern}})$ values within the matrix $\mathcal{C}_{\text{pattern}}$.
    
    \item \textbf{Establish Quantization Buckets:} Divide the range 
\[
[\min(\mathcal{C}_{\text{pattern}}), \max(\mathcal{C}_{\text{pattern}})]
\]
into $k$ equal intervals or 'buckets', each representing a quantization level.

    \item \textbf{Quantize Matrix Values:} For each element $c_{i,j,k}$ in the matrix $\mathcal{C}_{\text{pattern}}$, assign it to a quantization level based on which bucket its value falls into. This assignment is represented as a function $Q(c_{i,j,k})$ that maps the value of $c_{i,j,k}$ to one of the $k$ quantization levels.
\end{enumerate}

The output is a quantized 3-dimensional matrix where each element is represented by one of the $k$ quantization levels, effectively reducing the original range of values in $\mathcal{C}_{\text{pattern}}$ to $k$ distinct values.
\end{defn}

The $k$-Quantization of the matrix leads to generation of $k$ non-overlapping clusters of $\mathcal{C}_{\text{pattern}}$ and subsequently $\mathcal{C}_{\text{cons}}$. We use this non-overlapping partitioning of the matrix as a basis for sanitizing and releasing the electricity data $\mathcal{C}_{\text{cons}}$. Once partitioning is completed, the values in each partition are aggregated and sanitized based on the Laplace mechanism. The accumulated value in each partition is then uniformly distributed across its corresponding cells. More formally, denote the set of generated non-overlapping partitions by $\mathcal{P} = \{ P_1,\,P_2,...,P_k\}$ where each $P_i$ is a set of cells. Note that, partitions are generated based on $\mathcal{C}_{\text{pattern}}$ which is safe to release. The partitions are then used for matrix $\mathcal{C}_{\text{cons}}$, i.e. to compute the sanitized consumption values. A partition's cells are not necessarily continuous and may be scattered across the matrix. To sanitize and publish the electricity consumption values, the corresponding values in each partition are added and sanitized based on Laplace noise to achieve differential privacy, as follows:

\begin{equation}
    f(P_i) = \sum_{c\in P_i} f(c)  +Lap(s/\epsilon),
\end{equation}

where $c$ denotes a cell and the function $f(.)$ returns the added value of all cells in the partition. Once the sanitized value of each partition is generated, it is uniformly distributed among its cells. Therefore, for all $c\in P_i$ its value is updated to $f(P_i)/|P_i|$ in the sanitized matrix $\mathcal{C}_{\text{sanitized}}$.

A critical aspect is the allocation of privacy budget across quadtree levels. Theorem~\ref{theorem: partition sensitivity} establishes that the sensitivity of each partition is equal to the maximum number of cells contained within a single $xy$-axis {\em pillar} of the consumption matrix, where a pillar refers to all cells that have the same $x$ and $y$ coordinates. This theorem provides a foundational understanding of how sensitivity is distributed across the partitions, and guides the privacy budget allocation.

\begin{thm}\label{theorem: partition sensitivity}
    Let $P_i \in \mathcal{P}$ be a partition in the consumption matrix. The sensitivity of $P_i$ is the maximum number of cells it contains in any of the $xy$-axis pillars.
\end{thm}


Armed with this knowledge, the optimal assignment of privacy budget to each partition can be derived as follows. Let us denote the sensitivity of partition $P_i$ and allocated budget to this partition by $s_i$ and $\epsilon_{i}$, respectively. The optimal assignment of privacy budget to partitions can be formulated and solved by convex optimization as shown in Theorem~\ref{theorem: optimal privacy}.

\begin{thm}\label{theorem: optimal privacy}
    Given a non-overlapping partitioning of the consumption matrix \(\mathcal{P} = \{P_1, \ldots, P_m\}\) and the sensitivity of these partitions \(\mathcal{S} = \{s_1, \ldots, s_m\}\), the optimal allocation of the privacy budget to a partition \(P_i\) is derived by the following equation:
    \begin{equation}\label{equation: budget}
        \epsilon_i = \dfrac{\epsilon_{\text{sanitize}} \times s_i^{\frac{2}{3}}}{\sum_{i=1}^m s_i^{\frac{2}{3}}},
    \end{equation}
    where \(\epsilon_{\text{sanitize}}\) represents the total sanitization budget.
\end{thm}



\section{Experimental Evaluation}\label{section: experimental evaluation}



\subsection{Experimental Setup}\label{section: experimental setp}



{\em Datasets \& Spatial Distribution.} We conducted our experiments using four publicly accessible datasets, each under two distinct spatial distributions, resulting in a total of eight datasets. The statistics of these datasets are illustrated in Figure~\ref{figure: stat} and detailed in Table~\ref{tab:electricity-consumption} of the Appendix~\ref{appendix: table of datasets}. 
\begin{itemize}
    \item  CER~\cite{cer_smart_metering_project}: The dataset released by the Commission for Energy Regulation (CER) in Ireland originates from the Electricity Smart Metering Customer Behavior Trials carried out between 2009 and 2010. This project involved over 5,000 households and businesses and was focused on assessing the impact of smart meters on electricity consumption patterns. The objective was to gain insights for conducting a cost-benefit analysis regarding the country-wide adoption of smart meters. The anonymized data collected from these trials has been made accessible online for public research purposes.
    \item California, Michigan, and Texas Datasets~\cite{thorve2023high}: The datasets serve as digital twins representing residential energy usage within each state's residential sector. They are identified by the state's acronym and concentrate on the electricity consumption of the first five counties in alphabetical order for each state. For instance, the CA dataset includes data from Alameda, Alpine, Amador Butte, and Calaveras counties. These datasets provide hourly household electricity time series data from September to December $2014$. 
\end{itemize}


The privacy concerns regarding household-level electricity consumption have limited the availability of publicly accessible datasets, with no geotagged datasets currently accessible online~\cite{babaei2015study}. Therefore, to account for the distribution of users, we perform our experiments by distributing households in two settings: Uniform and Normal. 
A grid with granularity of $32\times 32$ is overlaid on the map, and the households are distributed over the space according to one of the two distributions. The center of the normal distribution is selected randomly over the map, and the households are located with the standard deviation equal to one-third of the grid size. The experiment is repeated $10$ times and the average result is shown to ensure repeatability of the experiments. Therefore, in total, the experiments are conducted on four datasets which are referred to as CA-Uniform, MI-Uniform, TX-Uniform, CER-Uniform, CA-Normal, MI-Normal, TX-Normal, and CER-Normal.\\

\begin{figure*}[t]
	\subfloat[CER; Random Shape \& Size Queries]{  
	\includegraphics[ scale = .3]{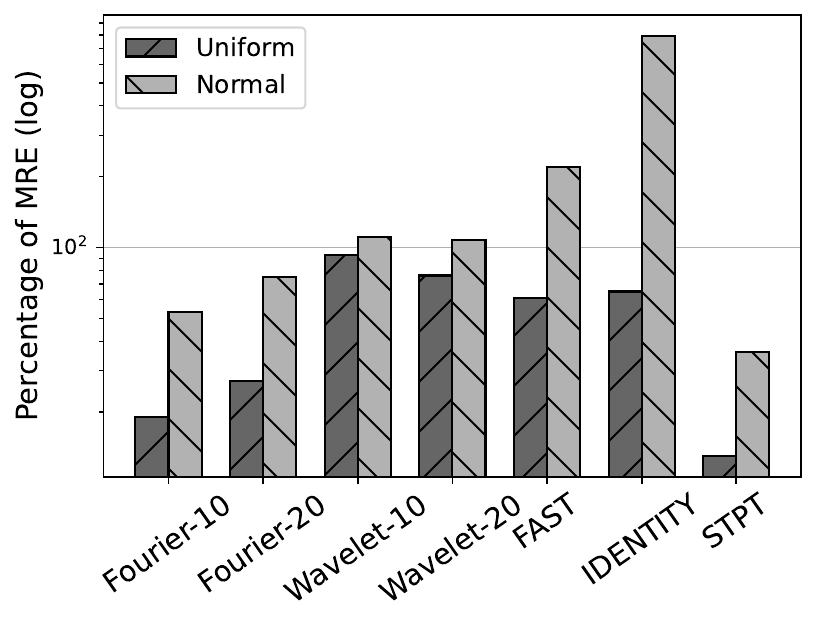}  
	}
	\hfill
	\subfloat[CER; Small Queries]{%
	\includegraphics[ scale = .28]{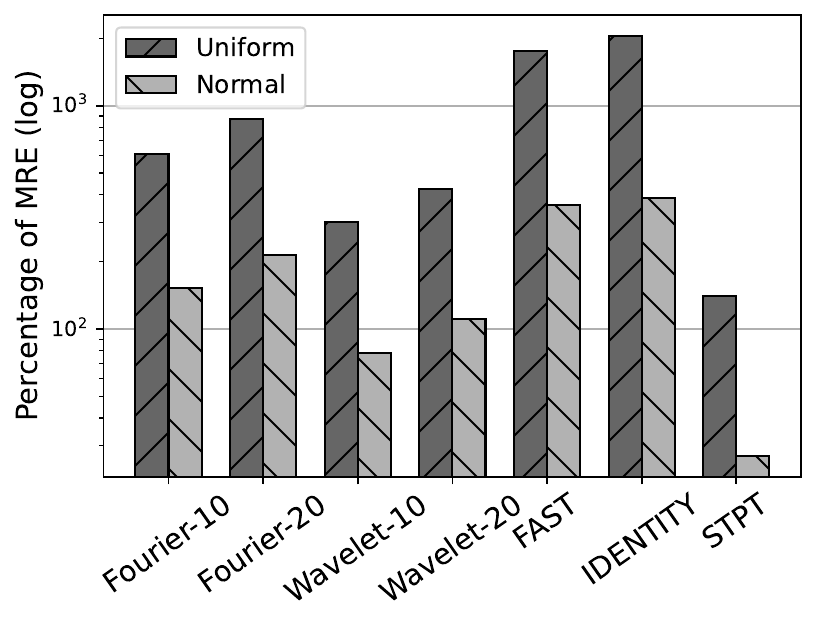}
	}
	\hfill
	\subfloat[CER; Large Queries]{%
	\includegraphics[ scale = .28]{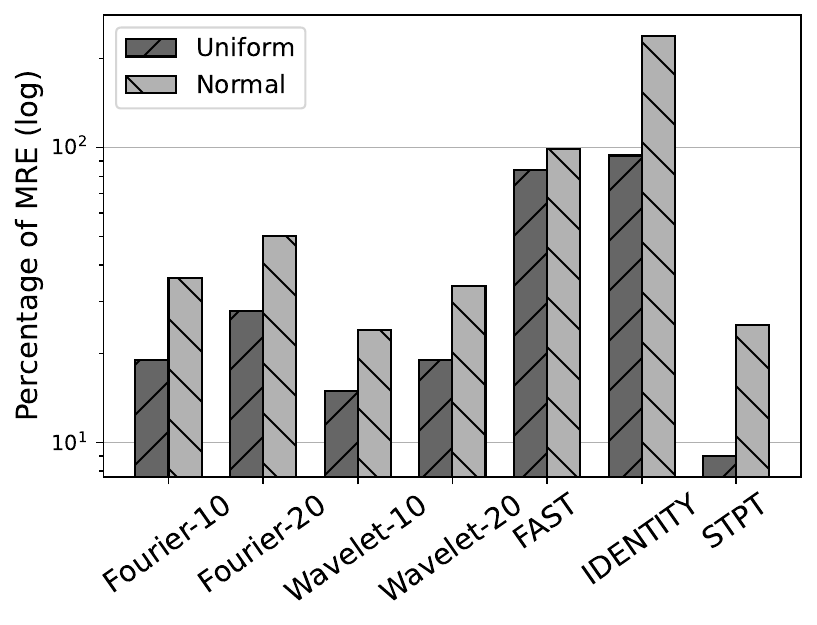}
	}
    \newline
	\subfloat[CA; Random Shape \& Size Queries]{  
	\includegraphics[ scale = .3]{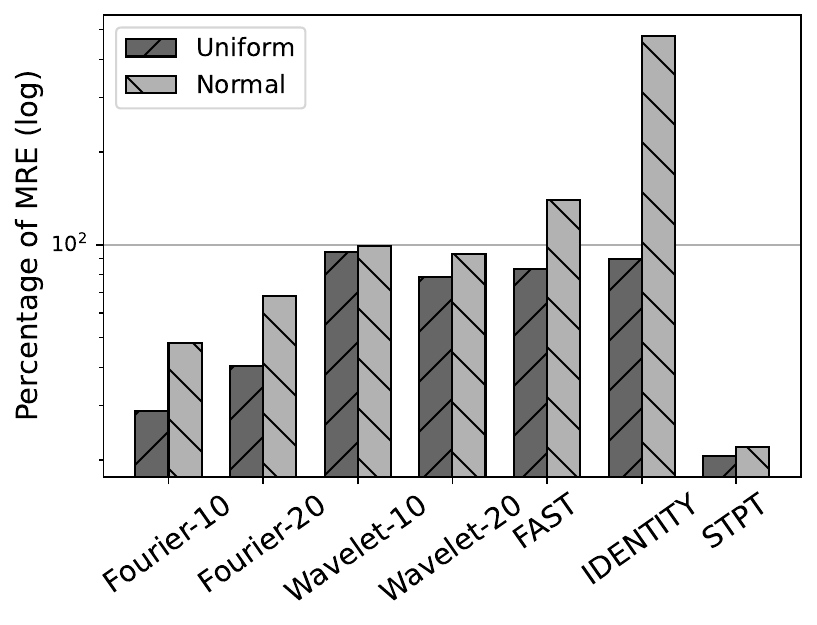}  
	}
	\hfill
	\subfloat[CA; Small Queries]{%
	\includegraphics[ scale = .28]{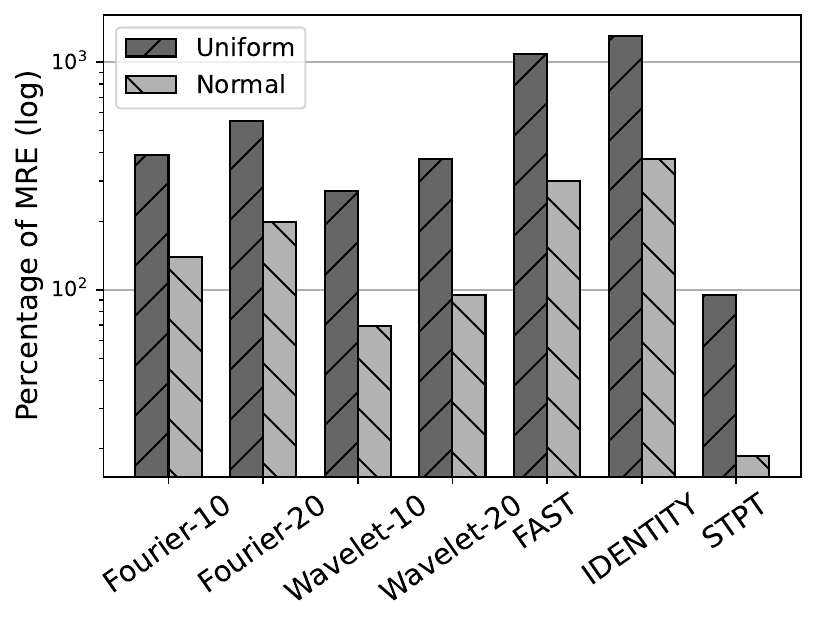}
	}
	\hfill
	\subfloat[CA; Large Queries]{%
	\includegraphics[ scale = .28]{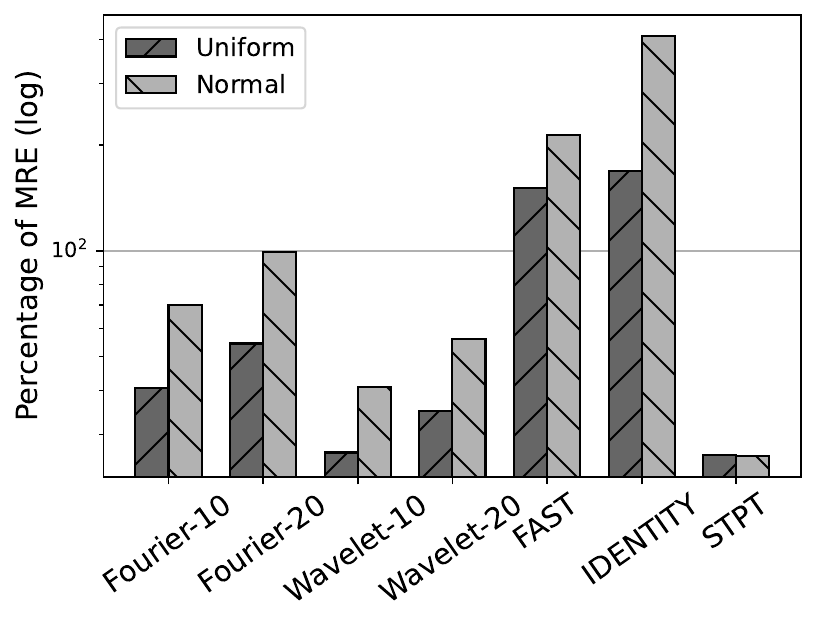}
	}
    \newline
	\subfloat[MI; Random Shape \& Size Queries]{  
	\includegraphics[ scale = .28]{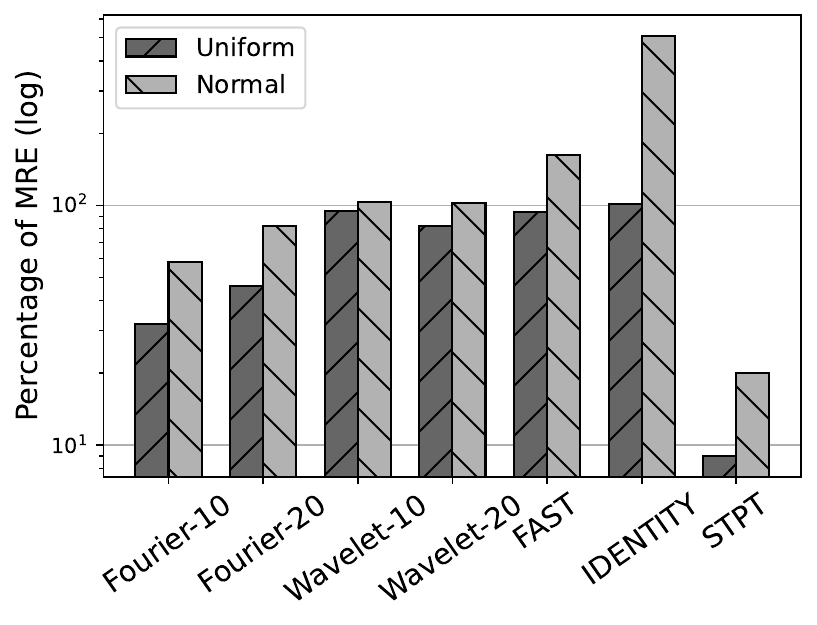}  
	}
	\hfill
	\subfloat[MI; Small Queries]{%
	\includegraphics[ scale = .28]{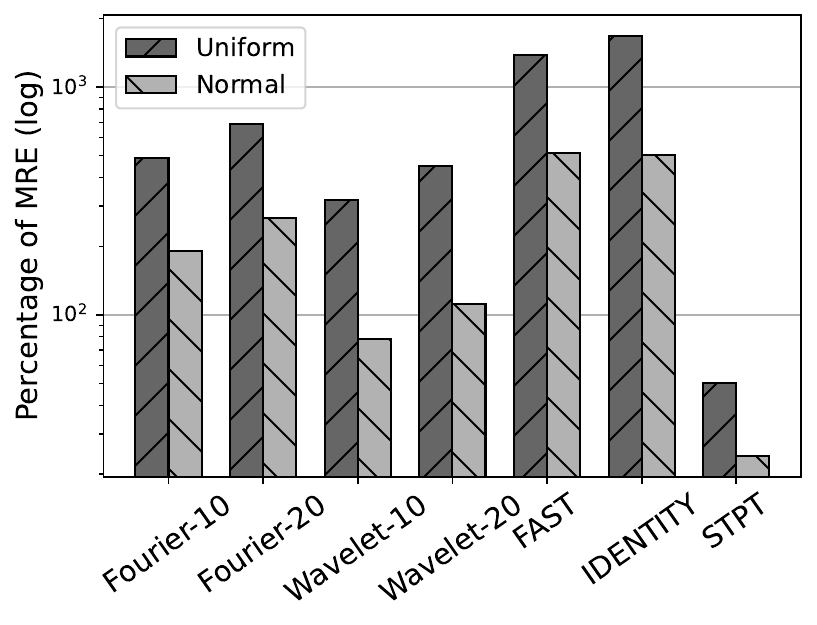}
	}
	\hfill
	\subfloat[MI; Large Queries]{%
	\includegraphics[ scale = .28]{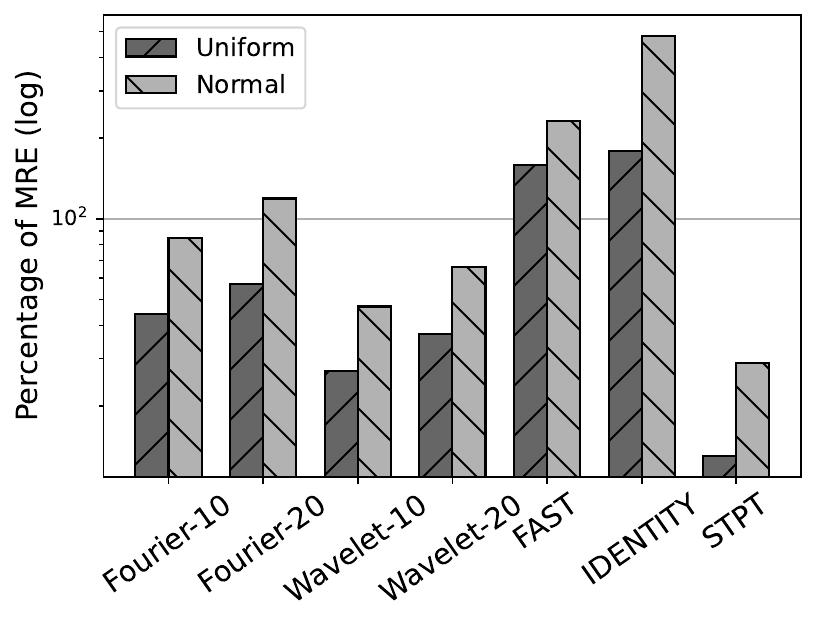}
	}
    \newline
	\subfloat[TX; Random Shape \& Size Queries]{  
	\includegraphics[ scale = .28]{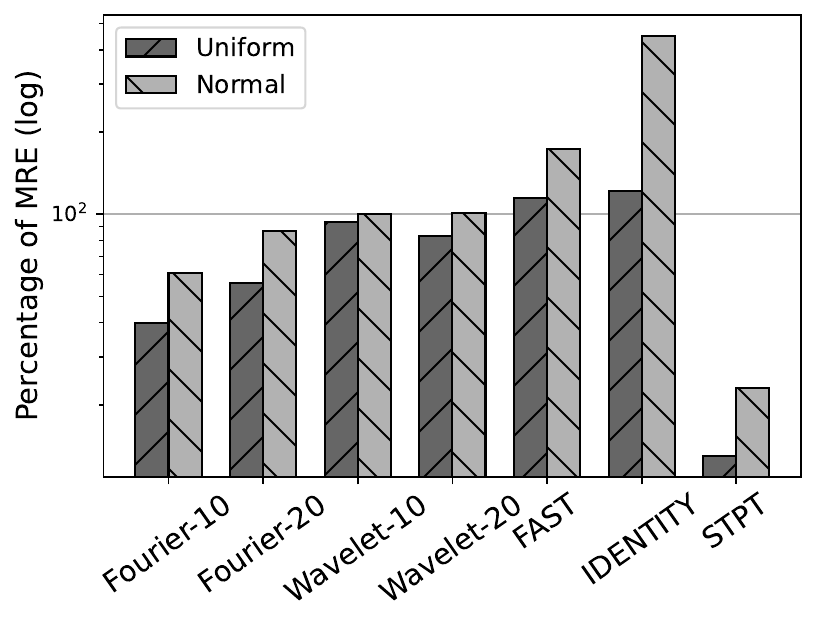}  
	}
	\hfill
	\subfloat[TX; Small Queries]{%
	\includegraphics[ scale = .28]{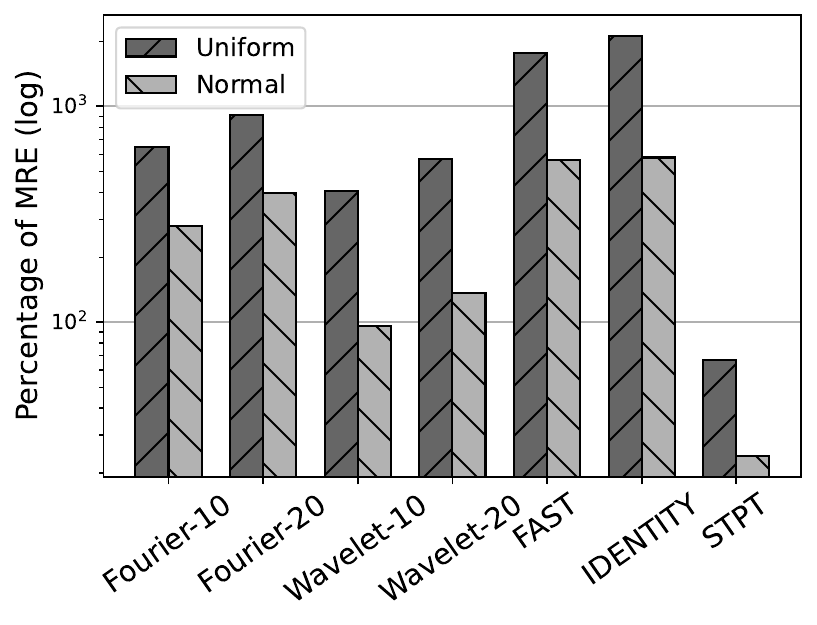}
	}
	\hfill
	\subfloat[TX; Large Queries]{%
	\includegraphics[ scale = .28]{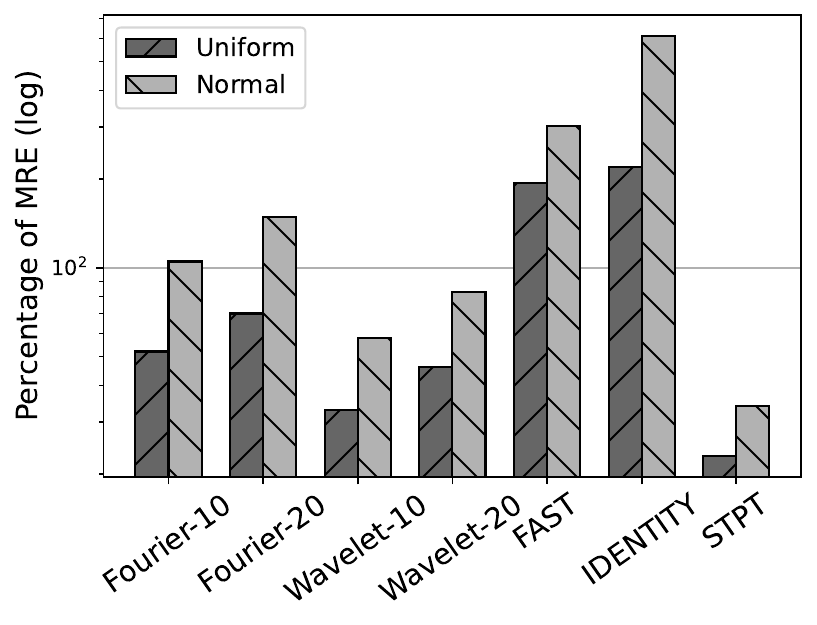}
	}
	\caption{Assessment of Algorithm Performance Across Datasets and Query Types. }
	\label{Fig: performance evaluation}
\end{figure*}

\noindent
{\em Benchmarks.} We compare the performance of our approach with the available state-of-the-art approaches detailed below. 

\begin{itemize}
    \item {\em FAST.} The framework proposed in~\cite{fan2013adaptive} is a widely adopted approach focused on exploiting the Kalman Filter for lowering utility loss while sanitizing time series. 
    \item {\em Fourier Perturbation Algorithm.} The methodology initially introduced in~\cite{rastogi2010differentially} and subsequently refined through sensitivity evaluations in~\cite{leukam2021privacy}, involves processing a time series with a specified integer $k$. The procedure begins by executing a Fourier transform on the time series, followed by the selection and sanitizing of the top $k$ primary Fourier coefficients. After sanitizing these coefficients, the inverse Fourier transform is applied, and DP time series are generated. We implement the algorithm in two settings where $k=10$ and $k=20$, denoted in the experiments as Fourier-10 and Fourier-20, respectively.
    \item {\em Wavelet Perturbation Algorithm.} By substituting the Fourier transform with the discrete Haar wavelet transform, Lyu et al. ~\cite{lyu2017privacy} introduced the wavelet perturbation algorithm for creating DP time series. This method, akin to the Fourier technique, requires an integer $k$, which signifies the number of coefficients to be used and sanitized. We denote this algorithm as Wavelet and implement it in two distinct scenarios: one with $k=10$ and the other with $k=20$.
    \item{\em Identity.} The approach illustrated in Subsection~\ref{section: simple strategy} is an adaptation of the original approach for the publication of time series and will be used as a comparison benchmark in our experiments. 
\end{itemize}

\noindent
{\em Query Types.} As discussed in the problem formulation of Section~\ref{section: Problem Formulation}, analysts are interested in range queries which are $3$-orthotopes with dimensions $d_1\times d_2 \times d_3$, indicating the consumption on a map region over a particular time range. For this purpose, we use small ($1\times 1\times 1$) and large queries ($10\times 10\times 10$) as well as queries with random shape and size. For each of the three categories, we generated $300$ randomly generated queries over the consumption matrix, calculated the MRE, and reported the average result.\\   

\noindent
{\em Hyper-parameters Setting.} Provided in Appendix~\ref{appendix: hyperparameters}.


\noindent
{\em Hardware and Software Setup.} Provided in Appendix~\ref{appendix: hardware}.

\subsection{Comparison with Benchmarks}\label{section: Performance Evaluation}

Figure~\ref{Fig: performance evaluation} illustrates the performance of algorithms when subjected to queries of differing shapes and sizes. Each row in the figure is dedicated to one of four datasets: CA-Uniform, CER-Uniform, CA-Normal, and CER-Normal. Within each row, the leftmost figure depicts the performance for randomly shaped and sized queries generated over the consumption matrix. The center figure shows results for smaller queries, and the rightmost figure displays the performance for larger queries. As can be seen, significant improvements have been made by STPT across the datasets in either distribution. As an example, for queries with random shapes and sizes, the STPT algorithm exhibited percentage-wise improvements of $60$, $31$, $54$, and $32$ in the Uniform setting for each respective dataset. Notably, the performance enhancement of the algorithms is more pronounced for smaller-sized queries. This result is desirable, indicating that more precise information about the consumption matrix can be conveyed with minimal loss of utility.

As anticipated, the IDENTITY algorithm generally shows the least accuracy among the baseline algorithms. However, it surpasses some of the more recent algorithms in scenarios where the data exhibit a more uniform shape, as seen in the first and second rows. An unexpected outcome of our experiments is the relative performance of Wavelet and Fourier transformations. Although Wavelet transformation was introduced at a later stage than the Fourier approach, the Fourier method demonstrates superior performance for queries of random shape and size. 

Another notable observation is that, on average, all algorithms tend to perform worse with non-uniform data. This aligns with findings in~\cite{shaham2021htf}, where a crucial determinant of performance is the homogeneity in data partitioning. Uniform data distribution contributes to higher homogeneity, and also decreases the uniformity error when estimating the size of random queries based on the sanitized partition counts.

\begin{figure*}[t]
	\subfloat[Impact of privacy budget on pattern recognition MAE.\label{figure: budget mae s1}]{  
	\includegraphics[ scale = .3]{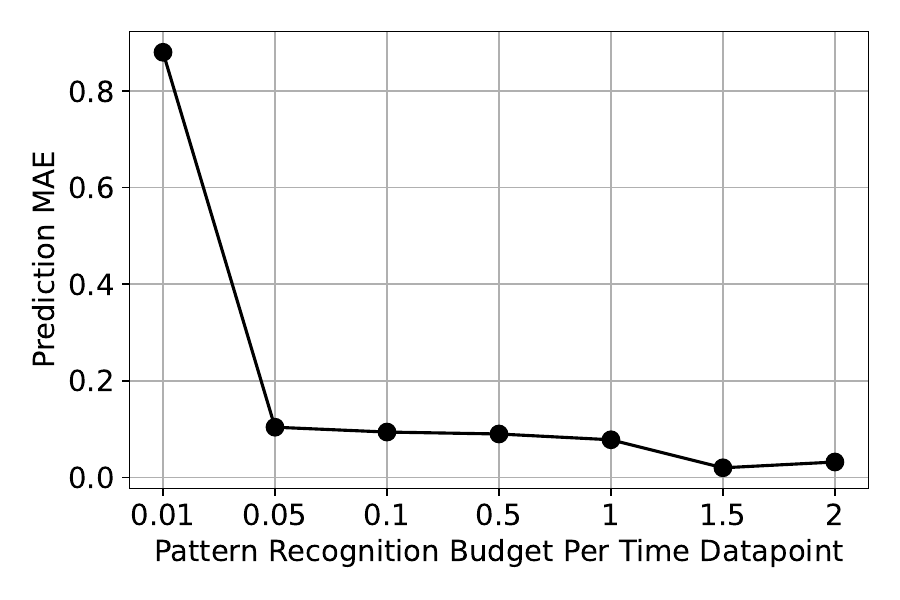}  
	}
	\hfill
	\subfloat[Impact of privacy budget on pattern recognition RMSE.\label{figure: budget rmse s2}]{%
	\includegraphics[ scale = .3]{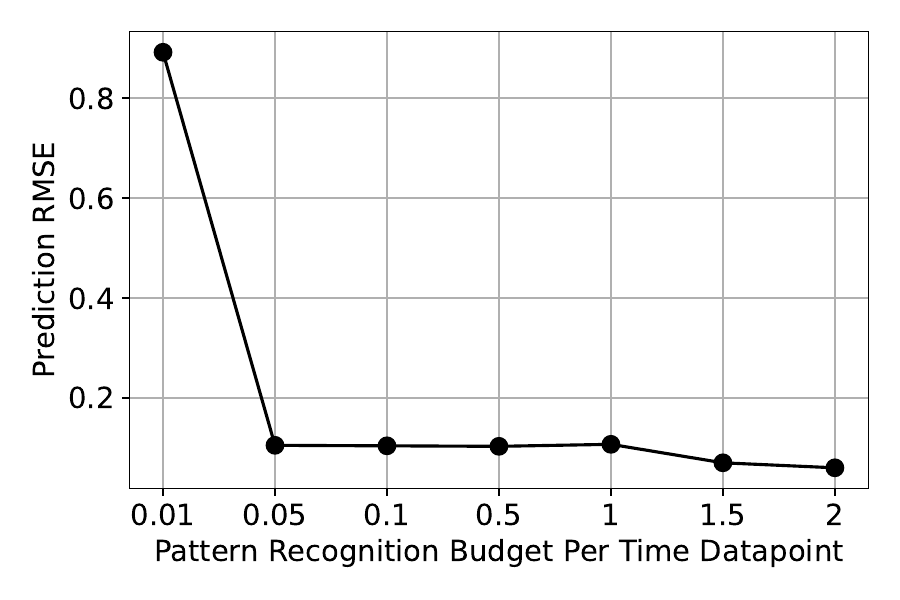}
	}
	\hfill
	\subfloat[Impact of quantization on performance.\label{figure: quantization s4}]{%
	\includegraphics[scale=.3]{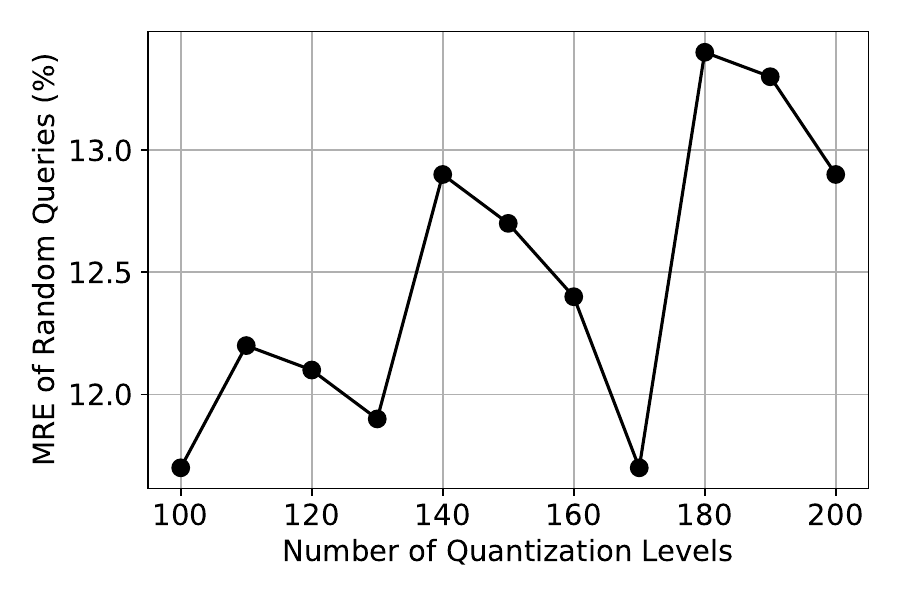}
	}
 \newline
	\subfloat[Computational complexity of algorithms.\label{figure: computational complexity s5}]{%
	\includegraphics[scale=.3]{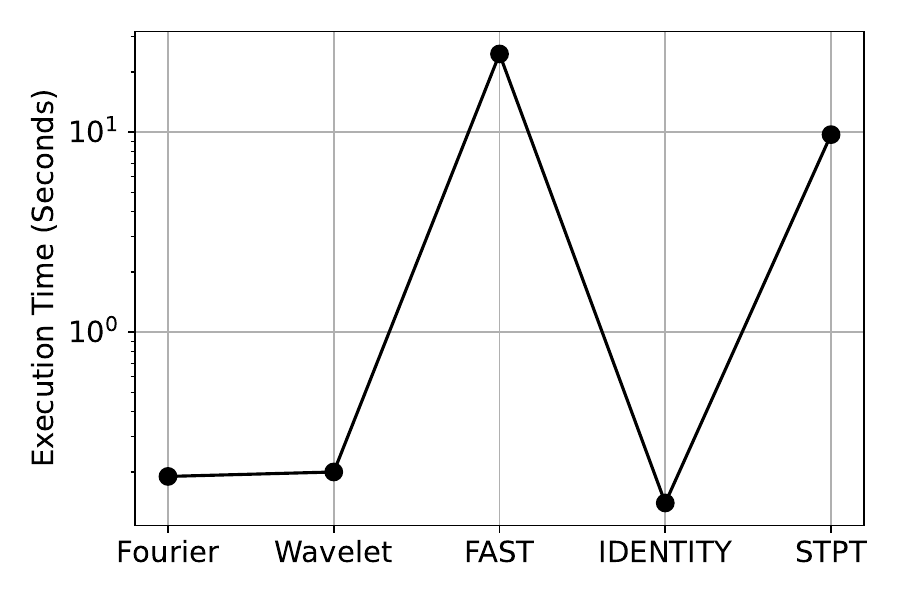}
	}
	\hfill
	\subfloat[Impact of Quad tree's depth on the prediction MAE of pattern recognition.\label{figure: quad depth mae s3}]{%
	\includegraphics[ scale = .3]{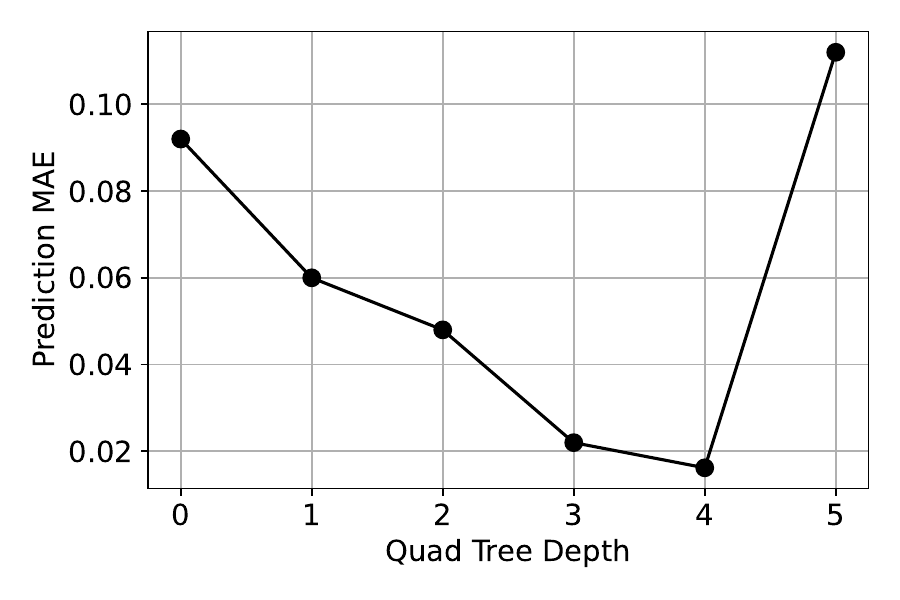}
	}
 \hfill
 	\subfloat[Impact of Quad tree's depth on the prediction RMSE of pattern recognition.\label{figure: quad depth rmse s4}]{%
	\includegraphics[scale=.3]{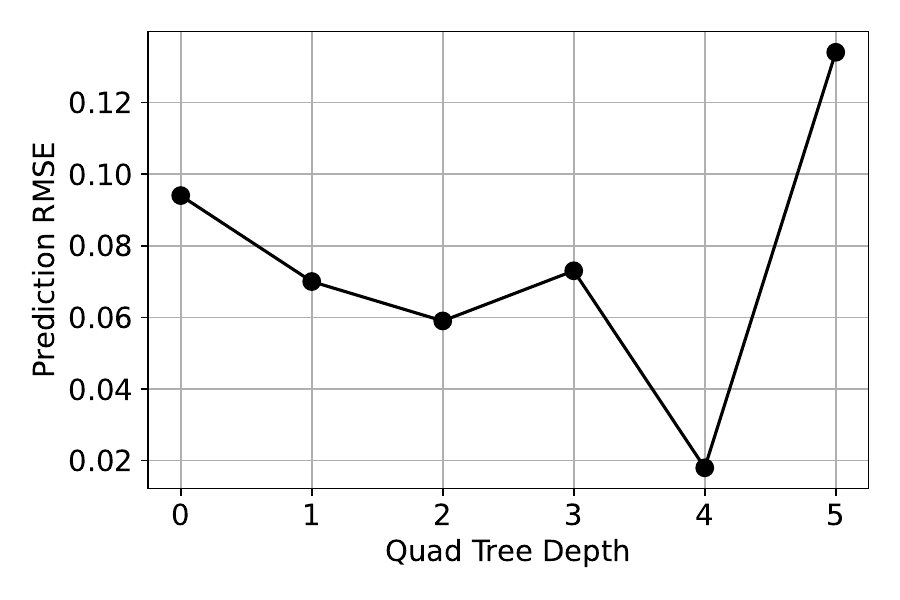}
	}
	\caption{Detaileld Analysis of STPT.}
    \label{figure: detailed analysis}
\end{figure*}

\subsection{STPT Detailed Evaluation}\label{section: Detailed Evaluation}


{\em Privacy Budget.} Figures~\ref{figure: budget mae s1} and~\ref{figure: budget rmse s2} analyze how the allocation of privacy budget affects pattern recognition performance in the STPT algorithm. While the sanitization budget in the second step remains constant, the budget for pattern recognition varies. For enhanced clarity, the $x$-axis displays the amount of budget allocated to each training datapoint of the RNN unit. The $y$-axis, meanwhile, indicates the MAE and RMSE of the RNN unit's predictions. As anticipated, an increase in the allocated budget enhances prediction accuracy, showcasing the privacy-utility trade-off. Notably, a significant improvement is observed when the privacy budget is increased from $0.01$ to $0.05$, suggesting that the minimal budget required for effective training lies within this range.

{\em Quantization.}  Figure~\ref{figure: quantization s4} illustrates the effect of the number of quantization levels on the performance of the STPT algorithm. The MRE metric is displayed on the $y$-axis for queries of varying shapes and sizes. Although there are fluctuations in the results, the general trend indicates that excessive increase in the number of quantization levels can negatively impact the effectiveness of STPT. This is expected, as many points in the cycle of a time series often exhibit similar values. Consequently, a high degree of quantization results in excessive partitioning and a reduction in the homogeneity that is captured in the data.

{\em Computational Complexity.} Figure~\ref{figure: computational complexity s5} presents and compares the runtime of various algorithms. According to the figure, the execution time for all algorithms is remarkably small, typically spanning just a few seconds. Although the STPT algorithm shows a slight rise in computational complexity, it is crucial to note that a significant portion of this complexity stems from the initial training phase required for pattern recognition, which is a one-time process. Overall, all algorithms demonstrate comparable execution times in the order of seconds, indicating that computational complexity does not pose a significant hurdle to their performance.

{\em Quad Tree Depth.} The influence of varying tree depth on pattern recognition efficiency is showcased in Figures~\ref{figure: quad depth mae s3} and~\ref{figure: quad depth rmse s4}. The aim is to explore how changes in tree depth affect the MAE and RMSE in the RNN unit. It's important to remember the balance between sensitivity and precision in time series produced at various depths. At shallow depths, such as the root node, the noise effect on relative error is low, since real counts are high. As the depth increases, this trend inverts. The results reveal that augmenting the tree depth up to a certain point enhances performance, but beyond that, the diminishing number of training data points at each level restricts further performance gains. Consequently, opting for a medium tree length, despite its impact on micro trends, proves to be advantageous.

\section{Related Work}\label{Sec: related work}


\noindent
{\bf Private Publication of Time Series.}
The existing body of work on differentially-private publication of time series falls into two primary categories: data transformation and correlation analysis. In the former category, the main strategy involves converting the data into an alternative domain that exhibits lower sensitivity, or provides a condensed representation of the time series. After sanitization in this new domain, an inverse function is used to revert the data back to its original form for publication. Notable methods in this category include the Fourier transformation~\cite{rastogi2010differentially,leukam2021privacy} and the discrete Haar wavelet transform~\cite{lyu2017privacy}. The latter category focuses on enhancing the utility of DP time series publications through improved leverage of inter-data correlations. This includes the concept of Pufferfish privacy, which employs a Bayesian Network to model correlations~\cite{song2017pufferfish}; the use of  Kalman Filters to reduce utility loss as explored in~\cite{fan2013adaptive}; and the adoption of a first-order auto-regressive process for correlation modeling as presented in~\cite{zhang2022differentially}.

\noindent
{\bf Private Publication of Multi-Dimensional Histograms.}
Our research closely aligns with the topic of publishing higher-dimensional histograms under DP. The study in~\cite{shaham2021htf} highlights the importance of data homogeneity in the private publication of histograms and introduces an algorithm called HTF (Homogeneous Tree Framework), designed to capture data homogeneity in order to reduce the effect of DP noise and thus improve utility. Another algorithm in this category is HDMM (High-Dimensional Matrix Mechanism)~\cite{mckenna2018optimizing}, which conceptualizes queries and data as vectors, and employs advanced optimization and inference methods for their resolution. DPCube~\cite{xiao2012dpcube} focuses on identifying and privately releasing dense sub-cubes. It allocates a portion of the privacy budget to derive noisy counts over a regular partitioning, which is subsequently refined into a standard kd-tree structure. The method then uses the remaining budget to acquire fresh noisy counts for the partitions, followed by an inference stage to rectify discrepancies between the two count sets. Other approaches, such as those in~\cite{AG} and~\cite{shaham2022differentially}, concentrate on modifying the granularity of space to enhance the utility of data in the publication of sanitized datasets.

\section{Conclusion}\label{section: conclusion}
Our study addressed critical privacy challenges in publishing electricity consumption data, balancing protection concerns with data utility. Our proposed innovative solution, STPT, significantly improves DP-compliant data publication accuracy by integrating time series data with the spatial attributes of households. This unique approach utilizes the short-term and long-term memory capabilities of RNNs for sophisticated pattern recognition, capturing both micro and macro consumption patterns. Our extensive experiments with real-world and synthetic datasets demonstrate STPT's superior performance in maintaining high data utility while ensuring robust privacy protection, compared to existing methods. The suggested method, while primarily aimed at enhancing the utility of DP compliant publishing electricity data, is versatile and can be applied to various situations, including Wireless Sensor Networks and the publication of health-related data, offering potential for future research applications.

\appendix

\section{Table of Notations}\label{appendix: table of notations}

Table~\ref{tab:table1} summarizes the notations used throughout the manuscript.

\newcommand{\rvec}{\mathrm {\mathbf {r}}} 

\begingroup
\begin{table}
\caption{Summary of Notations}
\centering
\begin{small}
\begin{tabular}{>{\arraybackslash}m{2.7cm} >{\arraybackslash}m{5.3cm} }
\hline\hline
  Symbol  & Description \\ \hline
    $\mathcal{D}$     & Time series data database \\
    $N$     & Number of households \\ 
    $\mathcal{U}$ & Set of households (or power grid users) \\
    $x_{i,t}$ & User $i$'s consumption at time $t$ \\
    $\mathcal{C}_{\text{cons}}$    & Actual consumption matrix \\
    $\mathcal{C}_{\text{norm}} $     & Normalized consumption matrix \\
    $\mathcal{C}_{\text{pattern}}$ & Pattern estimate matrix \\
    $\mathcal{C}_{\text{sanitized}}$ & Sanitized consumption matrix \\
    $\epsilon_{\text{tot}}$   & Total privacy budget \\
    $\epsilon_{\text{pattern}}$ & Pattern recognition budget \\
    $\epsilon_{\text{sanitize}}$ & Sanitization budget \\
    $\mathcal{P}$ & Partition set \\
    $s_i$ & Partition $i$ sensitivity \\
    $T_{\text{train}}$ & RNN training time \\
\hline\hline
\end{tabular}
\end{small}
\label{tab:table1}
\end{table}
\endgroup

\section{Proof of Theorems}

\subsection{Proof of Theorem~\ref{theorem: sensitivity simple strategy}}\label{Proof of Theorem 4}

    Recall that the consumption matrix is constructed such that the time series resolution matches the time axis resolution. As a result, each matrix cell contains no more than a single data point of an individual household/user. Consequently, adding or removing a user from the data can alter the value in a matrix cell by at most $\underset{i,t}{\max}\; x_{i,t}$. If the time series are normalized to values between $0$ and $1$, then this sensitivity would be $1$.

\subsection{Proof of Theorem~\ref{theorem: decomposition}}\label{Proof of Theorem 5}

The sequential decomposition in time is due to the correlation of time series over time. The parallel decomposition of the privacy budget over space is due to the fact that the time series of users are spatially bounded in the matrix and independent of the values in other cells.

\subsection{Proof of Theorem~\ref{theorem: sensitivity}}\label{Proof of Theorem 6}
Consider a cell at time $t$ corresponding to a sub-region at depth $i$ of the tree and all users $j$ falling in the sub-region. Let us denote the consumption of user $i$ before and after the removal of an individual by $x_{i,t}$ and $x_{i,t}'$, respectively. The maximum change observed in the representative time series of the sub-region denoted by $M$ at time $t$ can be derived as,
\begin{equation}
    \dfrac{ | \sum_{i\in M} x_{i,t} -   \sum_{i\in M} x_{i,t}' |}{4^{log_2(C_x)-i}} = \dfrac{ | x_{j,t} -  x_{j,t}' |}{4^{log_2(C_x)-i}} \leq \dfrac{ 1}{4^{log_2(C_x)-i}}
\end{equation}
In the above equation, index $j$ denotes the datapoint corresponding to the user whose existence in the dataset is altered. Therefore, the addition or removal of an individual can change the value of the representative point by at most $\dfrac{ 1}{4^{log_2(C_x)-i}}$.

\subsection{Proof of Theorem~\ref{theorem: partition sensitivity}}\label{Proof of Theorem 7}

    Denote by $s$ the maximum number of cells in a $xy$-axis pillar within the partition $P_i$. Given that the maximum cell count in each $xy$-axis pillar is bounded by $p$, and each pillar represents a unique time series, the addition or removal of an individual alters the cumulative values in the cluster by at most $p$. Hence, the sensitivity is characterized by this maximal change.

\subsection{Proof of Theorem~\ref{theorem: optimal privacy}}\label{Proof of Theorem 8}

The amount of noise added to each partition can be quantified using the variance of Laplace noise. Here, the goal is to distribute the privacy budget across partitions such that the total variance of applied noise is minimized. Equation~\ref{Equ: geometric progression} formulates this goal as a convex optimization problem. 

\begin{align}\label{Equ: geometric progression}
    \underset{\epsilon_1...\epsilon_m}{min}\;\;\;& \sum_{i\Equal1}^m s_i^2/\epsilon_i^2\\
    & \sum_{i\Equal1}^m \epsilon_i=\epsilon_{\text{sanitize}}, \;\; \epsilon_i>0 \; \;\forall i=1...m
\end{align}
Writing Karush-Kuhn-Tucker (KKT)~\cite{boyd2004convex} conditions, the optimal allocation of budget can be calculated as:
\begin{align}
    L(\epsilon_1,...,\epsilon_m,\lambda) &= \sum_{i\Equal1}^m s_i^2/\epsilon_i^2+ \lambda ( \sum_{i\Equal1}^m \epsilon_i- \epsilon_{\text{sanitize}})\\
    &\Rightarrow \dfrac{\partial L}{\partial\epsilon_i} = - \dfrac{2s_i^2}{\epsilon_i^3}+ \lambda =0\\
    &\Rightarrow \epsilon_i = \dfrac{2^{1/3} s_i^{2/3}}{\lambda^{1/3}},
\end{align}
Substituting $\epsilon_i$'s in the constraint equation, the optimal budget at the $i$-th level is derived as
\begin{equation}
    \epsilon_i = \dfrac{ \epsilon_{\text{sanitize}} \times s_i^{2/3}}{\sum_{i\Equal1}^m s_i^{2/3}}.
\end{equation}

\section{Dataset Statistics}\label{appendix: table of datasets}

\begin{figure}[t]
  \centering 
  \begin{minipage}{.5\textwidth} 
    \centering 
    \subfloat[CER\label{figure: dataset CER}]{%
      \includegraphics[scale=.3]{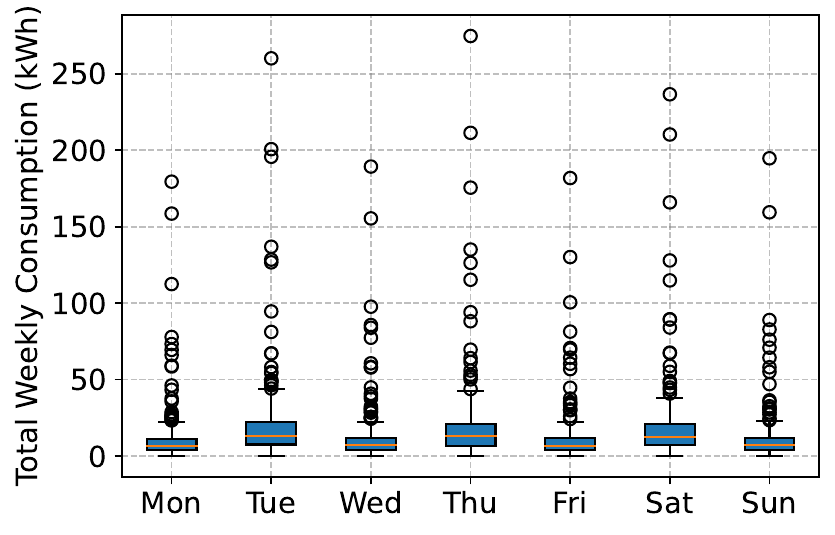}
    }
    \subfloat[CA\label{figure: dataset CA}]{%
      \includegraphics[scale=.3]{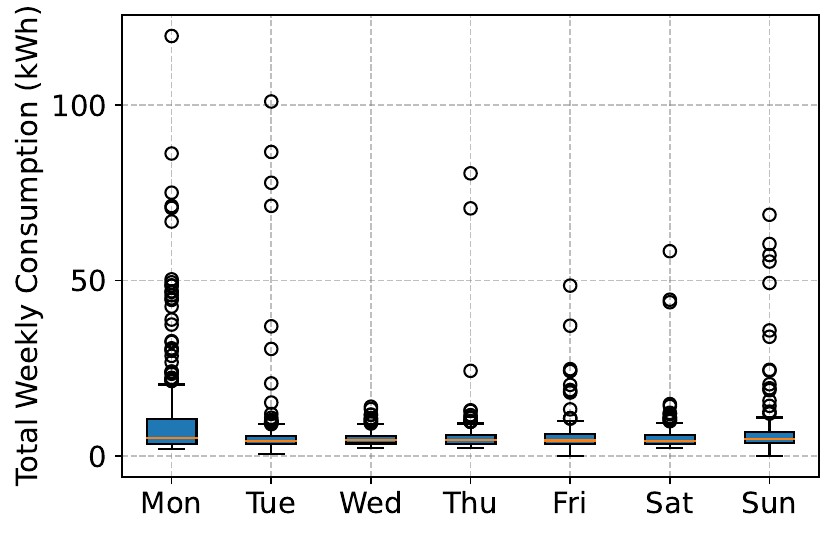}
    }
    \newline
    \subfloat[MI\label{figure: dataset MI}]{%
      \includegraphics[scale=.3]{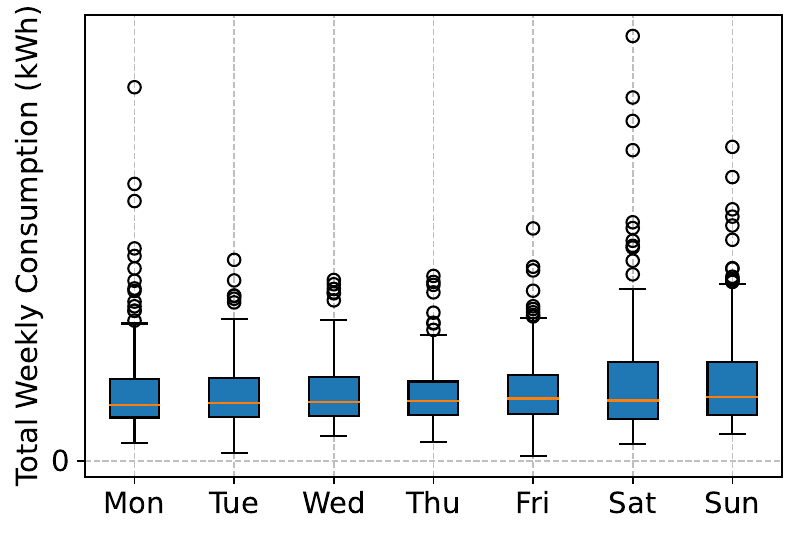}
    }
    \subfloat[TX\label{figure: dataset TX}]{%
      \includegraphics[scale=.3]{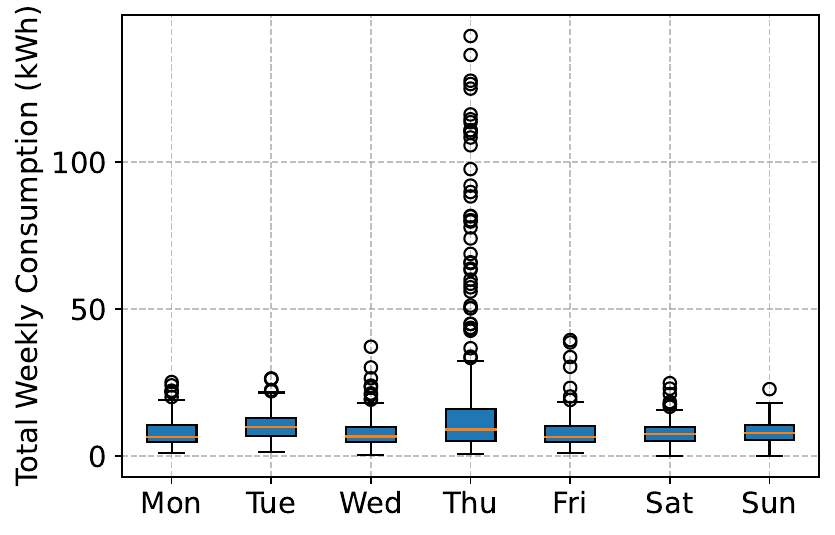}
    }
  \end{minipage}
  \caption{Total Weekly Consumption per Week Day.}
  \label{figure: stat}
\end{figure}

Table~\ref{tab:electricity-consumption} and Figure~\ref{figure: stat} summarize the statistics of datasets used in the experiments.

\begin{table*}
  \centering
  \caption{Electricity Consumption Data Summary}
  \label{tab:electricity-consumption}
\begin{small}
  \begin{tabular}{@{}lrrrrr@{}}
    \toprule
    Dataset & \multicolumn{1}{p{2cm}}{\raggedleft Number of Households} & 
              \multicolumn{1}{p{3cm}}{\raggedleft Average Hourly Consumption (kWh)} & 
              \multicolumn{1}{p{3cm}}{\raggedleft STD of Hourly Consumption (kWh)} & 
              \multicolumn{1}{p{3cm}}{\raggedleft Maximum Hourly Consumption (kWh)} & 
              \multicolumn{1}{p{3cm}}{\raggedleft Sensitivity Clipping Factor} \\
    \midrule
    CER & 5000 & 0.61 & 1.24 & 19.62 & 1.85 \\
    CA & 250 & 0.38 & 1.13 & 33.54 & 1.51 \\
    MI & 250 & 0.48 & 1.22 & 49.50 & 1.7 \\
    TX & 250 & 0.55 & 1.63 & 68.86 & 2.18 \\
    \bottomrule
  \end{tabular}
\end{small}
\end{table*}

\section{Hyper-parameters}\label{appendix: hyperparameters}

The total privacy budget is set to $\epsilon_{\text{tot}} = 30$, with $\epsilon_{\text{pattern}} = 10$ allocated for pattern recognition in STPT, and $\epsilon_{\text{sanitize}} = 20$ for sanitization. The same privacy budget is utilized across all algorithms. For training in the STPT algorithm, $100$ datapoints are used, resulting in a training matrix of $32\times 32\times 100$. The test involves $120$ points, leading to a matrix of $32\times 32\times 120$. Consequently, the published consumption matrix has dimensions of $32\times 32\times 120$. The sensitivity clipping factor of the consumption matrix is provided in Table~\ref{tab:electricity-consumption}. The RNN unit comprises a self-attention mechanism and a GRU unit. Training was conducted over $20$ epochs, with a batch size of $32$. The time window is set to encompass $6$ datapoints for predicting the next datapoint. The RMSProp optimizer is employed with a learning rate of 1e-3. The embedding size and hidden dimension are set to $128$ and $64$, respectively.\\ 

\section{Hardware and Software Specifications}\label{appendix: hardware}

Experiments were ran on a cluster node equipped with an $18$-core Intel i9-9980XE CPU, $125$ GB of memory, and two $11$ GB NVIDIA GeForce RTX 2080 Ti GPUs. Furthermore, all neural network models are implemented based on PyTorch version 1.13.0 with CUDA 11.7 using  Python version 3.10.8.

\newpage
\bibliographystyle{IEEEtran}
\bibliography{sample}

\balance

\end{document}